\documentclass[12pt]{iopart}
\pdfoutput=1
\usepackage{iopams}  
\usepackage{graphicx}
\usepackage{epsfig}
\usepackage{color}

\begin{document}

\title{Laser-plasma interactions for fast ignition}

\author{A J Kemp$^1$, F Fiuza$^{1,2}$, A Debayle$^3$, T Johzaki$^4$, W B Mori$^5$, P K Patel$^1$, Y Sentoku$^6$, L O Silva$^2$}

\address{$^1$Lawrence Livermore National Laboratory, CA, USA}
\address{$^2$GoLP/Instituto de Plasmas e Fus\~ao Nuclear - Laborat\'orio Associado, Instituto Superior T\'ecnico, Lisbon, Portugal}
\address{$^3$UPM, Madrid, Spain}
\address{$^4$University of Hiroshima, Hiroshima, Japan}
\address{$^5$Department of Physics \& Astronomy, University of California, Los Angeles, California 90095, USA}
\address{$^6$University of Nevada, Reno, USA}

\eads{\mailto{kemp7@llnl.gov}, \mailto{fiuza1@llnl.gov}}

\begin{abstract}
In the electron-driven fast-ignition approach to inertial confinement fusion, petawatt laser pulses are required to generate MeV electrons that deposit several tens of kilojoules in the compressed core of an imploded DT shell. We review recent progress in the understanding of intense laser plasma interactions (LPI) relevant to fast ignition. Increases in computational and modeling capabilities, as well as algorithmic developments have led to enhancement in our ability to perform multi-dimensional particle-in-cell (PIC) simulations of LPI at relevant scales. We discuss the physics of the interaction in terms of laser absorption fraction, the laser-generated electron spectra, divergence, and their temporal evolution. Scaling with irradiation conditions such as laser intensity are considered, as well as the dependence on plasma parameters. Different numerical modeling approaches and configurations are addressed, providing an overview of the modeling capabilities and limitations. In addition, we discuss the comparison of simulation results with experimental observables. In particular, we address the question of surrogacy of today's experiments for the full-scale fast ignition problem.
\end{abstract}

\maketitle
\normalsize

\section{Introduction}
\label{sec:intro}

In the fast ignition (FI) scheme \cite{Tabak94} of inertial confinement fusion (ICF), the compression and ignition phases are separated, offering the possibility for higher efficiencies with significantly relaxed symmetry requirements and target design constrains.

Ignition is triggered by a moderately short (10-20 ps) high-power (multi-PW) laser which hits the pre-compressed fuel, generating a population of fast electrons that carries a fraction of the laser energy to the core of the fuel, where energy is deposited through collisions, heating and igniting the fuel. To achieve ignition, $\sim 15-20$ kJ of energy must be deposited over a radius of $\sim 20 \mu$m in the dense core \cite{Atzeni99}. The electron beam energy required at the source strongly depends on the characteristics of the the fast electrons, namely its divergence and spectrum. In order to efficiently ignite the fuel the generated fast electrons must have low divergence \cite{Honrubia09} and energies within 1-3 MeV to ensure that they can reach the dense central core and be stopped \cite{Atzeni99,Atzeni04}. However, the control of the fast electron source is not trivial. It evolves the highly nonlinear absorption of an intense laser with a plasma density gradient ranging from sub-critical to overcritical densities, the generation of MA currents and giga-Gauss magnetic fields.

A considerable experimental effort has been carried in the last years to study the fast ignition interaction. The concept itself has been demonstrated in scaled-down experiments, showing efficiencies consistent with high-gain fusion \cite{Kodama01a,Kodama02}. The experiments of Kodama and Norreys have also introduced the cone-guided approach to fast ignition, as a way to open a corridor to the compressed core for the intense laser pulse and avoid the problems associated with laser plasma interactions in an underdense plasma, as anticipated in the original FI concept \cite{Tabak94}. It has also been shown that high-power lasers can efficiently transfer their energy to a population of relativistic electrons \cite{Green08,Ma12} and that these fast electrons can be transported and guided in high-density materials \cite{Kodama04,Green07}. However, important challenges remain to be fully understood including the dependence of the laser-generated fast electron characteristics (e.g. particle number, divergence, and temperature) on the laser/plasma parameters, and the dominant transport mechanisms in the high plasma density gradient. Furthermore, these studies are still far from ignition-scale conditions. 

The accurate description of the interaction of the intense ignition laser pulse with plasma, and the characterization of the laser-generated energetic electrons that emerge from the interaction region are essential elements of a point design for fast ignition.  The understanding of laser plasma interaction at intensities far above the relativistic threshold, pulse durations of thousands of laser periods and beam diameters of tens of laser wavelengths is an extreme challenging subject for experimental and computational science.

Lasers that access this parameter regime do not currently exist, which is why computational modeling has an important role in exploring the design space for fast ignition parameters.

Numerical simulations allow for a detailed understanding of the physical processes involved at these extreme conditions. However, due to the wide range of processes and scales involved, the modeling of relevant ignition conditions is extremely demanding. A hierarchy of numerical tools, from particle-in-cell (PIC) to hydrodynamic codes, is required to model the different scales of fast ignition, and an integrated modeling of the full interaction is not yet possible.

PIC codes \cite{Dawson83, Birdsall} are the main tool to resolve the spatial and temporal scales associated with the highly nonlinear and kinetic processes that occur during high intensity laser-plasma interactions. Therefore, PIC simulations are increasingly used in the study of FI. However, due to the need to resolve the small spatial and temporal scales associated with electron oscillations in the plasma, most simulations are of low dimensionality/scaled down system sizes. 
Limited by the cost associated with ignition-scale experiments and computer simulations, the last decade has seen a relatively small but growing number of studies devoted to this regime.

The scope of this article is to give an overview of current results, strategies, and methods that apply to fast ignition scale laser plasma interaction. We note that although in fast ignition electron generation and transport are intrinsically related, a separate review article will be focused on transport \cite{StrozziNF}. Here, we focus on the laser plasma interaction aspect of the fast ignition process.

The article is organized as follows. In Section \ref{sec:absorption}, after introducing the relevant laser- and plasma parameters, we begin by giving an overview of the main underlying physics mechanisms for absorption under FI relevant conditions. We then characterize a typical fast ignition electron source in Section \ref{sec:source}. We discuss the role of preformed plasma, in terms of the predominant acceleration mechanisms as well as in terms of numerical studies that have appeared in the recent literature. Preformed plasma (pre-plasma) can have a deleterious effect on the laser-to-electron coupling efficiency because absorption in the underdense plasma can lead to a much hotter spectrum compared to cases without low-density plasma. It also leads to a partial absorption of the laser pulse, which reduces the coupling efficiency into the 'useful' part of the electron spectrum.  We then discuss recent numerical studies of cone-guided FI and novel approaches like the double-wall cone, which promise a better energy coupling to the FI core by reducing the distance between the laser-plasma interaction and the core.

In Section \ref{sec:algorithms}, we discuss recent advances in numerical modeling of intense laser plasma interaction, and motivate their importance for a comprehensive understanding given current computational resources. 

Finally, we give an overview of current experimental results and an assessment of their surrogacy for fast ignition in Section \ref{sec:experiments} and we present our general conclusions in Section \ref{sec:conclusions}.

\section{Laser absorption and fast electron generation}
\label{sec:absorption}

\subsection{Regime of interest}
\label{sec:regime}

We begin by defining a parameter regime of interest for FI, for which we review the relevant physics models and computer simulations in the context of fast ignition-scale laser plasma interaction. The goal of FI is to deposit around 10-20 kJ of electron energy in the compressed core of an ICF capsule. The required energy at the source will ultimately depend on the electron beam properties such as the energy distribution and its divergence, as well as on the details of the target, in particular the distance between the laser absorption region and the dense core. Integrated simulations of electron transport in compressed-core configurations indicate that the required energy of the electron source is around 100 kJ or higher \cite{Strozzi12}. The deposition time should be around $10 - 20$ ps and is given by the hydrodynamic response of the heated material and its size \cite{Tabak94, Atzeni99}. The diameter over which this 10PW beam power is irradiated depends on the core diameter at peak compression; it is typically assumed that the electron beam diameter is between $30-100 \mu$m \cite{Strozzi12}. This means that the peak intensity of the electron beam is $10^{20}-10^{21}$ W/cm$^2$, and, assuming a coupling efficiency of 50\%, the resulting laser intensity is $>2\times10^{20}$ W/cm$^2$. For these parameters a simple 'ponderomotive' scaling argument gives typical electron energies greater than 5MeV, which is undesirable because of their excessive electron stopping length in the core plasma at density of 200-600g/cm$^3$ and a diameter of around 100$\mu$m. It could be advantageous to go to shorter laser wavelengths in order to reduce the electron energy to the optimum stopping range around 1-3MeV. Cost and efficiency considerations with respect to current laser technology, however, limit the laser wavelength to values between 0.3-1 $\mu$m. Laser intensity and -wavelength are connected through a dimension-less parameter that determines the equations of motion \cite{Sarachik70}. The normalized laser amplitude $a_0=eE_L/m_e\omega_L c$ is related to the laser electric field $E_L$ and its wavelength $2\pi c/\omega_L$; a laser intensity of $I_0=1.37\times10^{18}$ W/cm$^2$ at 1$\mu$m wavelength corresponds to $a_0=1$. For fast ignition, the laser normalized vector potential is in the range $1 \ll a_0 < 30$, which is called the 'ultra-intense' or 'relativistic' interaction regime, because a single electron interacting with such a laser in vacuum will oscillate with energies $\frac{a_0^2}{2}m_e c^2$, which significantly exceed their rest mass, $m_e c^2$. At low intensities the electron oscillates primarily in the transverse direction, however,  at relativistic intensities the electronÕs momentum in the laser propagation direction, $\frac{a_0^2}{2}m_e c$ begins to exceed its momentum in the transverse direction, $a_0m_e c$. As a result, as discussed in the next section, when a laser interacts with an overdense plasma electrons are rapidly accelerated forward into the plasma and the energy is found to scale as the so called ponderomotive energy \cite{Wilks92}
\begin{equation}
E_p=m_ec^2(\sqrt{1+a_0^2}-1)
\end{equation}
during a laser cycle, for a linearly polarized laser pulse. 

\subsection{Energetic electron production mechanisms}
\label{sec:acc}

In the interaction of an ultra-intense laser with a fast ignition target several mechanisms are responsible for the absorption of the laser and generation of energetic electrons. These can occur in different regions of the laser-plasma interaction and also at different times, since the self-consistent evolution of the system will modify the plasma conditions in time;  and they can occur simultaneously.  

In this section we discuss the dominant acceleration mechanisms. It is only the  electric field that can perform work on the particles, so what differentiates the mechanisms is the source of the electric field (the laser or the plasma surface) and the distance over which the electric field does work. As we noted earlier, a plane wave laser in vacuum can accelerate an electron starting from rest to an energy of $\frac{a_0^2}{2}m_e c^2$. However, when a laser is reflected and absorbed at an over-dense plasma there is standing wave set up and there is a component of the light pressure from the standing wave at twice the laser frequency that creates a longitudinal electric field. In addition, the surface can ripple so that the angle of incidence of the laser at the surface is not normal and there is thus a component of the laser electric field that is normal to the surface. Determining which electric field is responsible for the energetic electron production is therefore complicated. The use of particle tracking in simulations is now helping to differentiate between the possible mechanisms.

Depending on the amount of sub-critical density plasma, acceleration can occur predominantly at (1) at a steep density gradient; (2) near dense plasma, with a small, i.e., a few laser wavelengths long, near-critical density-shelf of preformed plasma; and/or (3) over many laser-wavelengths worth of sub-critical density plasma. The critical density, i.e. the density at which light (at non-relativistic intensity) is reflected is defined as $n_c=m_e\omega_L^2/4\pi e^2$. Under realistic conditions, acceleration can occur in these different regions at the same time, which complicates the physics analysis of full-scale simulations and experiments. 

\begin{figure}[t!]
\begin{center}
\includegraphics[width=1.\textwidth]{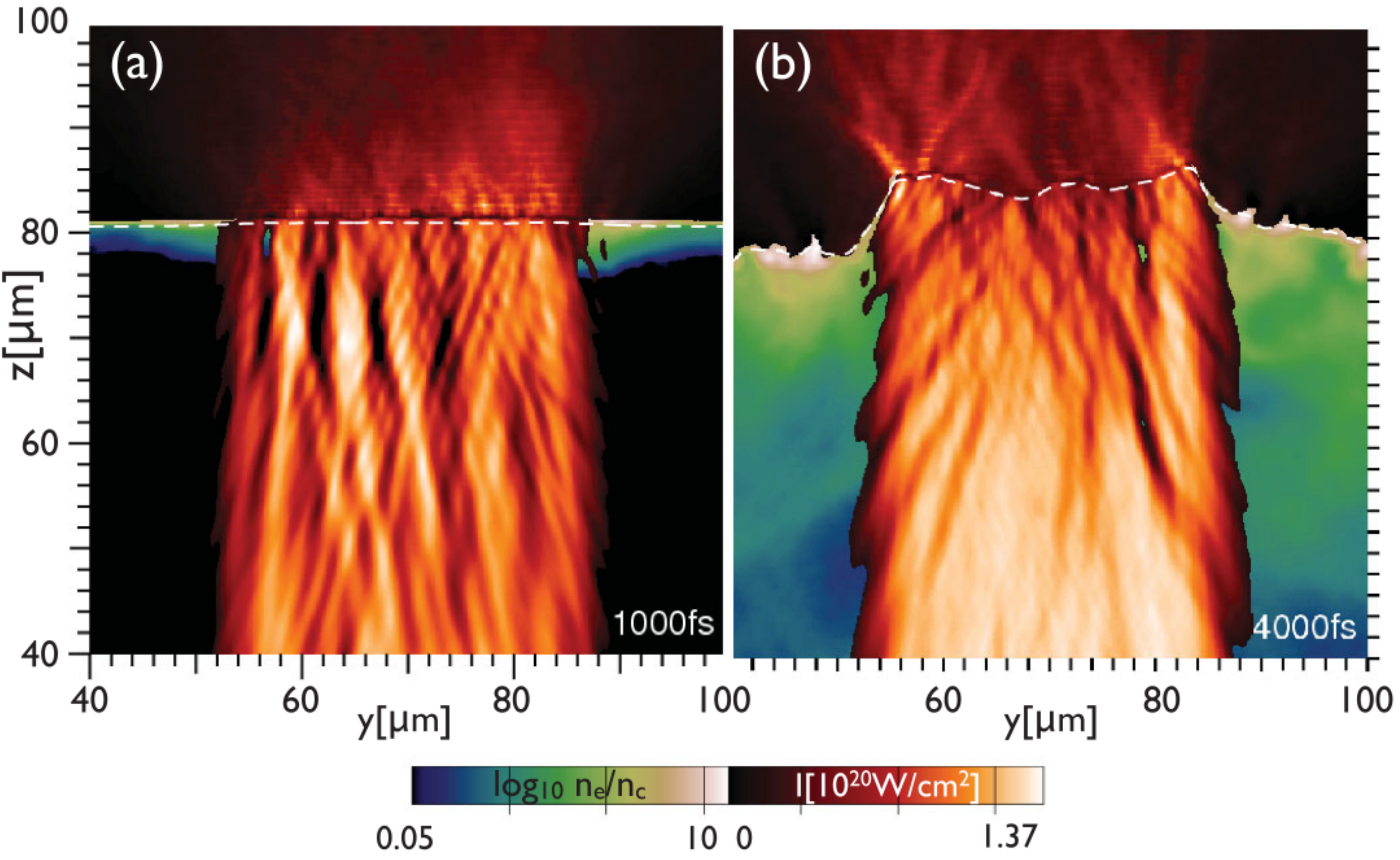}
\caption{\label{fig:lpi} Relativistic petawatt laser pulse interacting with over-dense plasma at 1 ps (a) and at 4 ps (b); the laser pulse is injected at $z = 0$, and plasma is initially at $z > 80 \mu$m. Energy flux density along z (in red) shows continuously high conversion from the laser into a relativistic electron beam. The dashed line at $n_e = 10n_c$ shows deformation and motion of the absorption layer. Expansion of under-dense plasma into vacuum (in green) is evident.}
\end{center}
\end{figure}

In the study of the laser absorption and fast electron acceleration it is common to start with a steep (or short scale length) plasma-vacuum interface. While there are hydrodynamic simulations of fast ignition-scale implosions available, they lack spatial and temporal resolution and detail at the critical density surface at the time when the short pulse laser would interact, i.e., 'realistic' plasma density profiles are uncertain. Hence we start our discussion with a generic representation of the preformed plasma. The conceptually most simple, and optimistic, approach is to set up simulations of PW laser interaction without any preformed plasma, using a uniform electron density of $\sim 100 n_c$, which is sufficiently over-critical to laser intensities corresponding to $a_0\leq30$, and small enough to allow for resolving the plasma skin depth at reasonable numerical cost (see Sec. \ref{sec:algorithms} for a discussion on the numerical limitations). Figure~\ref{fig:lpi} illustrates a typical set-up of a PIC simulation of intense laser plasma interaction at near-ignition scale~\cite{Kemp12}. In the beginning, the laser pulse interacts with a sharp density gradient, eventually the plasma will gradually expand, leading to an under-dense plasma that fills the vacuum in front of the bulk plasma. As the plasma evolves in time, different acceleration mechanisms will overlap and influence each other, e.g., the formation of a low-density plasma in front of the target leads to an increased laser-plasma interaction at low density, effectively reducing the power available for the first two mechanisms; and strong absorption in near-critical density plasma causes plasma heating, and thus an expansion, so that low-density plasma fills the volume in front of the critical surface. Note that because of the finite laser spot size $r_0$, the expansion remains quasi one-dimensional until its scale length exceeds the size of the laser spot, i.e., at $r_0/c_s\approx2$\,ps \cite{Kemp12}, where $c_s$ is the sound velocity, which explains the asymptotic scale length of under-dense plasma being similar to the laser spot diameter observed in our simulation. Figure 1(b) shows that the outer edge of the laser beam is evolving due to whole-beam self-focusing, so that the conditions at the edge over a length $\approx\lambda_L$ are different from the interior. As the relative volume of edge filaments is $2\lambda_L/r_0$, their role is small for a wide beam. We limit our discussion to fully ionized plasma, which has typically an ion-to-electron mass ratio of $2m_p/m_e$ nearly independent of material, and ignore atomic physics effects like bremsstrahlung losses or ionization. First, we will discuss the main acceleration mechanisms in an idealized set-up, i.e., starting from a plasma density profile that rises rapidly from vacuum to $100n_c$. 

\subsection*{Acceleration next to a steep density gradient}
\label{sec:steep}

A linearly polarized plane electromagnetic wave that is normally incident on over-critical density plasma is almost completely reflected back, leading to a standing wave in the vacuum half-space $z<0$ in front of the plasma. In the limit of negligible skin depth, the fields vanish on the plasma surface and the corresponding field pattern is  \cite{May11}
\begin{equation}
\hat{E}_y=2a_0\sin(k_z)\sin(\omega_L t)\\
\hat{B}_x=2a_0\cos(k_z)\cos(\omega_L t)
\label{eqn:standing}
\end{equation}
and $E_y=B_x=0$ for $z>0$. 
In order for electrons to gain significant energy from the laser, they must escape the plasma and enter the vacuum in order to experience the strong laser field and be accelerated. Electrons from the bulk plasma can only reach the peak of the electric field associated with the standing wave (a quarter wavelength from the surface) if they have finite momentum in the direction of the laser electric field polarization when they leave the plasma into the vacuum region. Electrons can then be rotated by the magnetic field at the plasma-vacuum interface so that they propagate in vacuum perpendicularly to the plasma surface allowing them to reach the location where the transverse electric field is at a anti-node, i.e., 1/4 wavelength away from the surface. These electrons will gain the maximum attainable momentum of $2a_0$ before being turned back into the target by the magnetic Lorentz force, leading to a characteristic cut-off in the energy distribution. Electrons leaving the plasma without a significant transverse momentum will simply be rotated back to the plasma by the magnetic field at the interface without gaining much energy.

\begin{figure}[t!]
\begin{center}
\includegraphics[width=1.\textwidth]{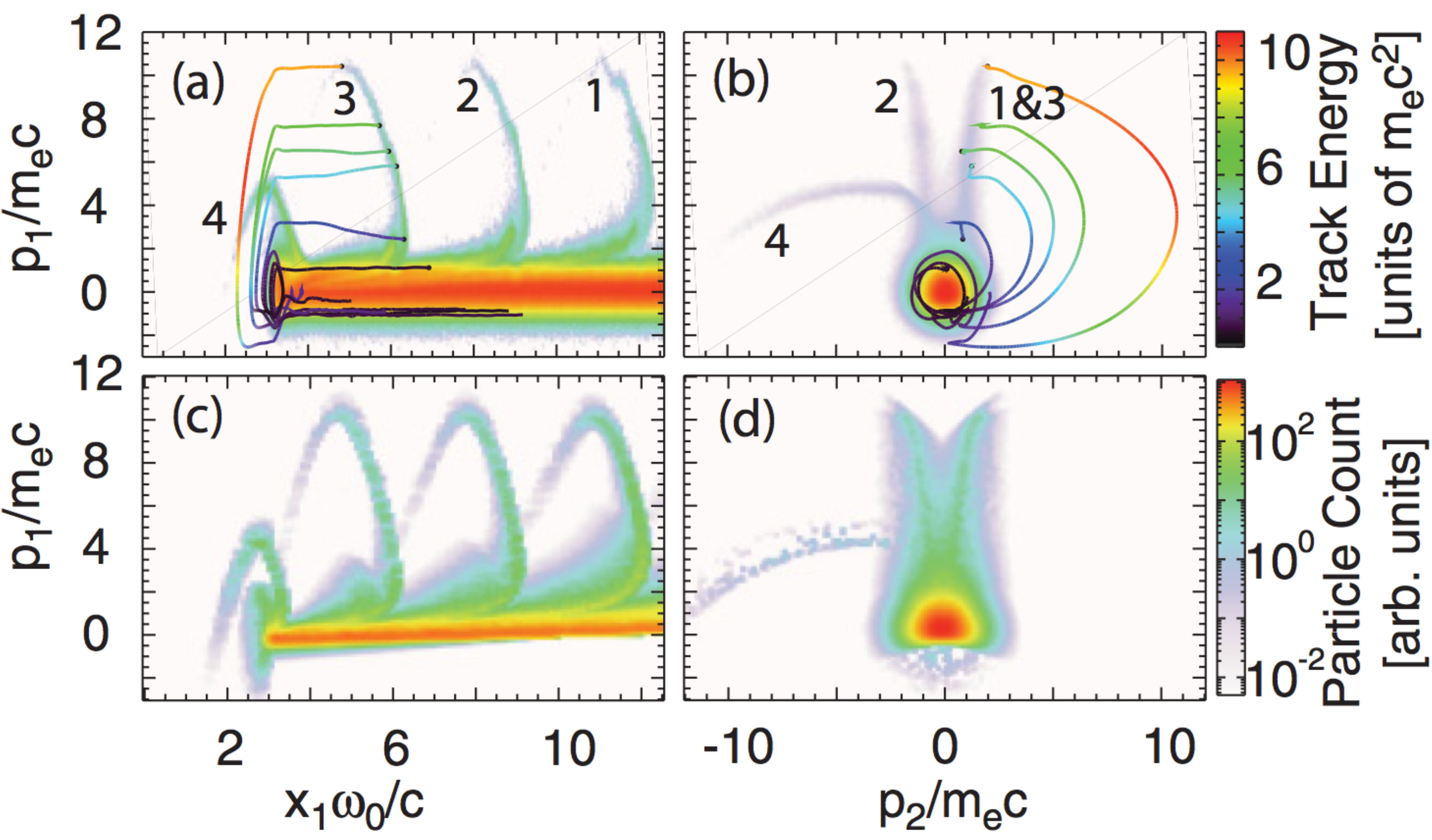}
\caption{\label{fig:2a0} Top row: electron phase spaces, (a) $p_1$ vs $x_1$ and (b) $p_1$ vs $p_2$, for an OSIRIS \cite{Fonseca02,Fonseca08} simulation with $a_0 = 6$ at $t=13.85 \omega_L^{-1}$, with tracks for individual particles superimposed. Bunches of electrons are labeled 1-4. Bottom row: Same phase space plots for test particles moving in a standing wave with $a_0 = 6$.}
\end{center}
\end{figure}

Figure \ref{fig:2a0} (a,b) shows a 2D OSIRIS \cite{Fonseca02, Fonseca08} PIC simulation of this process, illustrating the trajectories of accelerated particles in momentum space \cite{May11}. In this simulation the ions are immobile and the laser which propagates in the rise time is nearly instantaneous. In figure \ref{fig:2a0} (a,b) the snapshots are taken at a time of $\sim$ two laser cycles.  Due to the standing wave pattern Eq. \ref{eqn:standing}, we find two bunches of hot electrons per laser cycle, similar to what the $J \times B$ heating effect inside the relativistic skin-layer would give \cite{Kruer85}. The main differences are that (a) here the EM fields do not need to enter the skin layer to accelerate particles (acceleration occurs in vacuum \cite{Bauer07}) and (b) the fluid description on which the $J \times B$ model is based does not apply outside the plasma. This illustrates that the generation of bunches at $2\omega_L$ can occur from a variety of processes each of which are in response to the $v \times B$ force.

Figures \ref{fig:2a0} (c,d) correspond to a test particle simulation by May et al. \cite{May11} where test electrons propagate in the standing wave pattern in front of a plasma Eq. \ref{eqn:standing}. For a high initial transverse electron temperature near the plasma surface, such that the Debye length is comparable to the plasma skin depth, the qualitative agreement with the full PIC simulation results shown in Fig. \ref{fig:2a0} (a,b) confirms that at early times in the interaction, collective effects on the plasma surface are negligible. 

The results shown in Fig. \ref{fig:2a0} assume that the plasma has a thermal distribution with a temperature of 75keV, so that a small but constant amount of electrons emerges from the target without laser interaction. But even if the plasma electrons were initially cold, i.e., if the electron thermal velocity $v_{th}\ll c$, they would still be heated up from the laser plasma interaction, as will be discussed below.

The mechanism described here does not lead to significant absorption and energetic particle production of circularly polarized light, since the magnetic field is at an anti-node in the plasma-vacuum interface and does not oscillate with time but instead rotates, thereby preventing plasma electrons to escape the plasma into vacuum \cite{May11}. Under quasi-1D conditions, absorption will be only a few percent of the laser power, and mostly due to the static contribution to the ponderomotive force inside the skin layer \cite{Mishra09, Sanz12, Debayle13}.

As a result of the electron acceleration out of the thermal background, the ratio of transverse- to longitudinal particle momenta is smaller for higher longitudinal momenta energies, leading to the generation of electrons with a very low intrinsic divergence. Figure \ref{fig:2a0} (b) illustrates the phase space of the laser generated electron beam perpendicular to the laser propagation direction. It is important to note that although these electrons are accelerated in vacuum with low divergence they can still be affected by the fields inside the plasma, which can significantly increase their divergence. In addition, if the plasma self-consistently heats to MeV temperatures then the particles in the 1-3 MeV range will also have a large intrinsic angular divergence. 

\subsection*{Electron injection into the accelerating structure}
\label{sec:injection}

The recent work discussed previously has shed light on the details on how energetic electrons are produced in somewhat idealized circumstances. This work indicates that these electrons are generated by the interaction of electrons outside a steepened overdense plasma interface by a standing wave once the plasma is sufficiently heated.  This work shows that in order for electrons to escape outside the steep interface they need to be sufficiently energetic to escape the magnetic field at the surface. However, under realistic cases where the laser intensity gradually rises, the plasma self-heats, and the surface ripples the manner in which electrons escape into the ÒvacuumÓ region may be more complicated. These issues have also been discussed in the recent literature.. 

In a more gradual gradient or at lower intensities the penetration of the laser field in the finite plasma skin-depth region \cite{Yang95, Rozmus90, Catto77, Bauer07} can heat the plasma. The evanescent wave is able to naturally heat up the electrons in the skin layer to multi 10s keV. This would, after some delay time related to electron thermal velocity and laser pulse rise time, lead to the extraction of electrons out of the skin layer and acceleration in the standing wave into the plasma \cite{Bauer07, May11}. This is related to the initial $j \times B$ heating mechanism of Kruer and Estabrook \cite{Kruer85}.

Another possibility, is associated with the electrostatic field at the plasma-vacuum interface that arises at higher intensities and lower plasma temperatures where the excursion of an electron oscillating at $2\omega_L$ component of the light pressure in the standing wave is greater than the Debye length. In this case electrons can be heated by the associated longitudinal electric field. Sanz and Debayle et al \cite{Sanz12, Debayle13} have put forward a 1D model with immobile ions to describe the electrostatic field contribution on laser absorption. At the high laser intensities of interest for fast ignition, their model describes the motion of the electron plasma boundary induced by the laser ponderomotive force, which has a main component at $2 \omega_L$ and a small component at the plasma frequency. They propose that electron acceleration is mainly attributed to the oscillating piston formed by the standingwave and the electrostatic field, moving with the electron plasma boundary velocity. The electrostatic field inside the plasma oscillates at plasma frequency around a mean value equal to the laser ponderomotive force $-a_y(z,t) B_x(z,t)/\sqrt{1+a_y^2}$, where $a_y$ is the laser potential vector and $B_x$ is the laser magnetic field. The resulting total force inside the plasma oscillates around zero, and causes various bunches of electrons to be pushed into the standing wave, where they are accelerated back into the plasma by the increasing total force. In this mechanism, although the longitudinal work $j_z E_z$ is negligible compare to the transverse work $j_y E_y$, the electrostatic field is strong enough to expel electrons into the standing wave region with a characteristic energy well above the thermal energy. However, this method produces less energetic electrons since the electrons do not escape sufficiently into vacuum in order to experience the maximum electric field \cite{May11}. It is worth pointing out that this scenario of electron heating is efficient as long as the electron plasma boundary oscillation amplitude is comparable with the plasma skin depth.

Most theoretical models for absorption are 1D and for immobile ions. Capturing the initial laser absorption and injection into the standing wave structure requires multi-dimensional simulations with mobile ions. Furthermore, it should be noted that as the plasma surface becomes modulated the laser can penetrate this plasma ripples and heat up electrons through additional mechanisms, such as Brunel heating. May et al \cite{May11} showed that in multi-dimensional mobile ion simulations that an initially cold plasma  naturally heat up to 10s or 100s of keV at early times near the surface, independently of the details of the heating process, allowing for injection of electrons into the standing wave. In addition to the multi-dimensional aspect of the absorption, May et al. also show that the work done by the longitudinal electrostatic field is negligible for the electron acceleration (it is the laser transverse field that produces the acceleration).

\subsection*{Magnetic fields on surface}
\label{sec:bfields}

\begin{figure}[t!]
\begin{center}
\includegraphics[width=1.\textwidth]{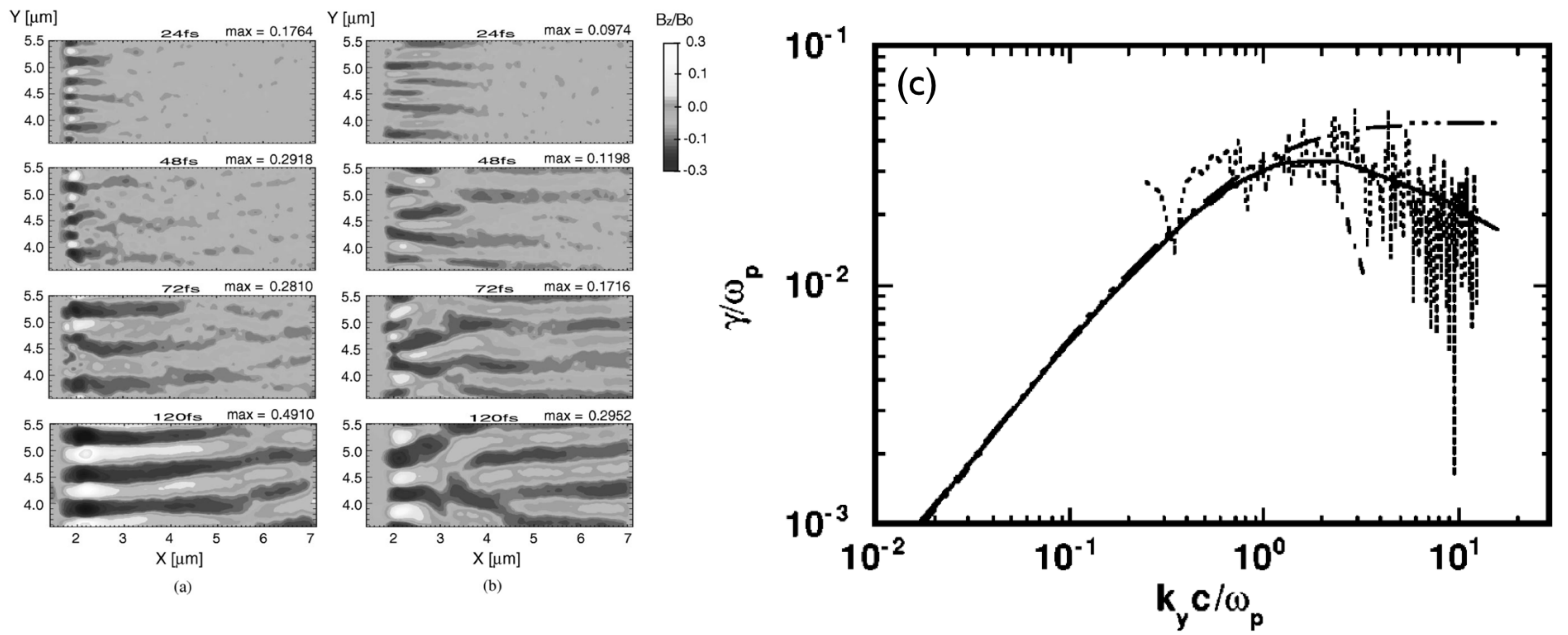}
\caption{\label{fig:weibel} (left) Temporal evolution of quasistatic magnetic field in $20 n_c$ plasma during irradiation with an $a_0 = 3$ laser pulse; (right) Growth rate of filamentation instability vs transverse wave number in simulation (dotted line) and linear analysis (solid line).}
\end{center}
\end{figure}

The directional electrons accelerated at early times will be subject to a filamentation instability \cite{Lee73, Honda00, Sentoku00, Taguchi01, Silva02} that sets in on a time scale of tens of plasma periods, i.e., after a few laser cycles for $n_e=100n_c$ plasma \cite{Sentoku00, Silva02, Ren04, Adam06, Tonge09}.  Over this time, magnetic field structures are formed near the plasma surface with amplitudes close to the laser magnetic field. Sentoku et al \cite{Sentoku00} have demonstrated in 2D PIC simulations and linear analysis of the beam filamentation instability that the growth rate peaks at a spatial frequency $k\simeq\omega_p/c$, compare Fig. \ref{fig:weibel}.

The magnetic fields related to particle beam-filaments close to the plasma surface can deflect electrons into a beam with finite opening angle. Scattering of fast electrons in the high amplitude magnetic fields will lead to an increase of the beam divergence and will smear out the features shown in Fig \ref{fig:2a0} at late times ($\simeq$ 100 fs). Adam et al \cite{Adam06} report that in 2D and 3D simulations after around 50fs, the electron beam fans out into a cone with an opening half-angle of around 20$^\circ$, defined for particle energies above 1MeV. This early-time electron beam divergence appears even before the plasma surface is perturbed significantly by the laser interaction. 

\subsection*{Instability of the plasma interface}
\label{sec:surface}

\begin{figure}[t!]
\begin{center}
\includegraphics[width=1.\textwidth]{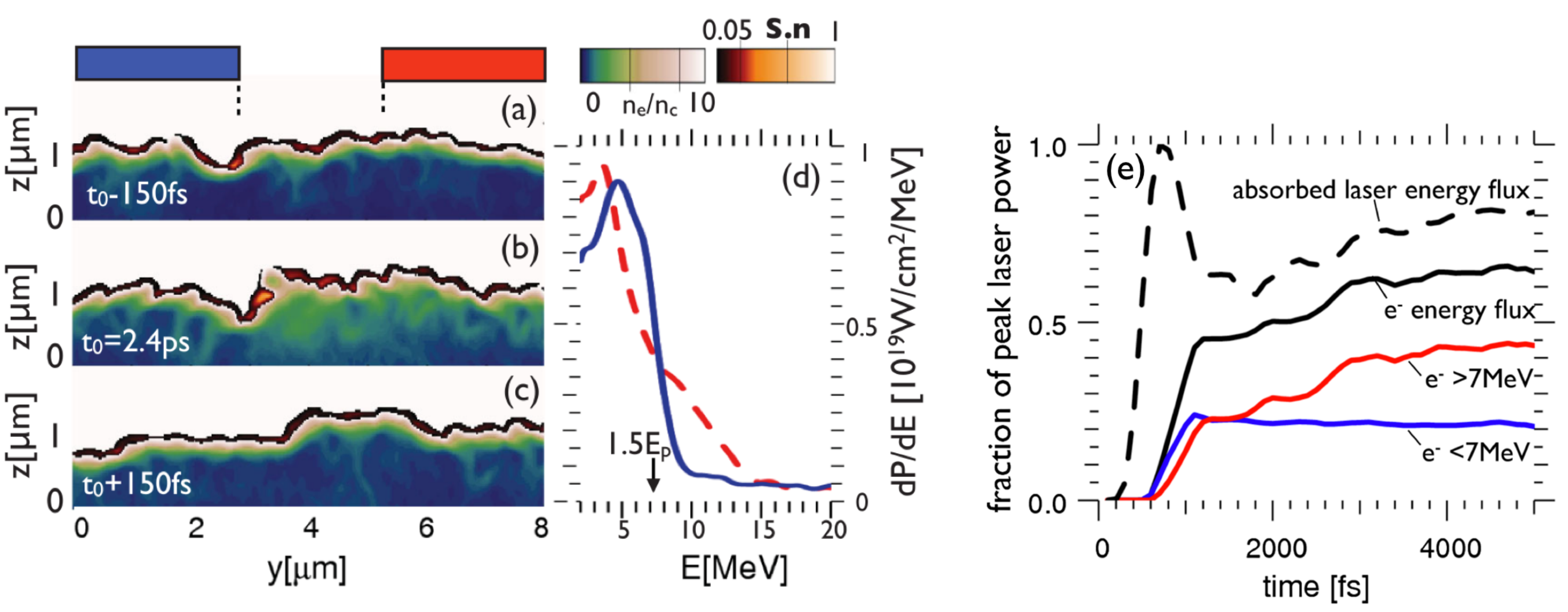}
\caption{\label{fig:spectrum} (left) Nonlinear saturation stage of plasma surface rippling driven by the laser interaction: (a)-(c) snapshots of laser Poynting flux normal to the $n=10n_c$ target surface in red and electron density in green; (d) electron energy spectra at 2.4ps determined in boxes with and without prior emission of plasma, as indicated by boxes on top of (a); (right) Time history of energy partition in laser generated electrons, showing sustained absorption of up to 80\% (absorbed laser energy flux through z = 0 plane, dashed line) into relativistic electrons (total electron energy flux projected on z, solid black line); also shown are contributions from particles with energies $E_{kin}\le1.5 E_p=7$ MeV and $>7$ MeV; all values are normalized to peak laser power $P_L = 1.3$ PW.}
\end{center}
\end{figure}

As the intense laser interacts with the steep plasma profile for multiple 100fs intervals, it will also lead to a modulation of the plasma interface. The dense plasma interface is unstable to transverse modes so that the absorption layer becomes porous, as shown in Fig. \ref{fig:spectrum} \cite{Kemp12}; the cause of this rippling is related to surface waves, filamentary, Rayleigh-Taylor like, and modulational instabilities \cite{Sagdeev59, Valeo75, Nishimura80, Gamaly93, Macchi02, Ren04, Tonge09, AnderBrugge12, Kemp12}. As the surface gets modulated the laser is able to directly accelerate electrons since linearly polarized laser light will have an electric field component along the target normal direction. The evanescent electric field inside the skin layer is able to periodically remove electrons into vacuum, so that absorption increases like $1/\mathrm{cos}(\theta)$. This effect has been originally referred to as 'not-so-resonant, resonant absorption', Brunel effect, or vacuum heating \cite{Brunel87}. It also leads to a larger number of hot electrons that escape from the plasma into vacuum and create a sub-critical plasma region over time. Surface modulations are, of course, absent in one-dimensional PIC simulations, which predict a steepening of the density gradient as well as reduced absorption \cite{Kemp09}. Only for very large density gradients it is possible to observe sustained absorption from under-dense plasma over two picoseconds \cite{Cai10}. 

Once the interaction surface becomes significantly rippled, the distinction between various absorption mechanisms becomes more complex, and the initial laser polarization does not play a significant role for absorption. Three-dimensional simulations \cite{Fiuza11, Kemp12} with linear polarization show that the laser-generated electron beam is relatively isotropic in the plane perpendicular to the laser propagation direction after a few 100 fs. Corresponding 2D simulations (where isotropy along one direction is assumed), on the other hand, give results that depend strongly on the laser polarization. Over several hundred femtoseconds the absorption is high only if the laser magnetic field is aligned with that associated with the electron beam filamentation. Furthermore, close to the absorption region the difference between linear- and circular polarization becomes obsolete. After a period of a few hundred femtoseconds, the electron beam observed in 3D simulations with circularly polarized light resembles that found in the case of linearly polarized light. 

We note that as the surface gets rippled, the strong electric and magnetic fields that are generated around the surface will have an important impact on the trajectory of individual laser-accelerated electrons, and larger divergence angles are observed. We have performed test-particle simulations in which a narrow beam of 'infinitesimal-charge' electrons is injected into a region with a field structure taken from a snapshot of a fully self-consistent PIC simulation. The test particles exhibit the same divergent behavior as that found in the PIC simulation, i.e., the initially narrow beam diverges immediately after entering the region with strong electric and magnetic fields and then propagates in ballistic fashion inside the dense plasma \cite{Fiuza12c}.

The release of plasma through the porous interface allows for a near-constant recession of the absorption layer along the laser direction. At an intensity of $1.4\times 10^{20}$W/cm$^2$ and at a density of $100n_c$, i.e. the parameters of the simulation shown in Fig. \ref{fig:spectrum} \cite{Kemp12} we find a recession velocity of $v_s=5\times 10^{-3}c$ along the laser direction. This value can be derived from momentum and energy flux conservation between the laser on one side, and the plasma electrons and ions on the other side. We write energy and momentum flux balance along the laser irradiation axis $z$ between laser and plasma components at the absorption plane $z_0$ as
\begin{eqnarray}
(I_+ + I_-)/c&=&P^{(e)}+P^{(i)}\\
I_+ - I_-&=&F^{(e)}+F^{(i)}
\end{eqnarray}
where $I_{+/-}$ denote incident and reflected intensity, $P_{e,i}$ the momentum-, and $F_{e,i}$ is the energy flux density of electrons or ions along the laser irradiation axis $z$. The latter quantities are defined by
\begin{eqnarray}
\label{eq:Pei}P^{(e,i)}_{z}&=&\int_p f^{(e,i)}(p,\,z_0)\,p_z\,\beta_z{\rm d}^3p\\
F^{(e,i)}_{z}&=&\int_p f^{(e,i)}(p,\,z_0)\,(\gamma(p)-1)\,m_ec^2\,\beta_z{\rm d}^3p\ ,
\end{eqnarray}
where $\beta\equiv v/c$ and $\gamma\equiv\sqrt{1+p^2}$, and $f_{e,i}$ are the electron- and ion distribution functions, respectively.

Due to symmetry around the laser axis, lateral energy and momentum fluxes cancel out so that we can focus on the forward-going components. For relativistic laser intensities, the electrons' momentum and energy distributions peak at relativistic energies where $\gamma\gg1$ and $\beta\approx c$ so that $F^{(e)}\approx c\,P^{(e)}$, while the ion energy flux $F^{(i)}$ is negligible. Therefore the momentum transferred to plasma ions is $P^{(i)}\simeq2\,I_-/c$, or, in other words, ion momentum is mostly gained from the elastically backscattered light. Together with the expressions $I_+\lambda^2/c=a_0^2n_c m_e c^2/2$ for the incident light, $I_-=(1-f_a)I_+$ for the reflected light and $P^{i}= 2v_s^2\,n_i\,M_i$ for the ion momentum flux density, we arrive at the relation
\begin{equation}
2(1 - f_a)\,a_0^2\,n_c\,m_e\,c^2/2 = 2M_i\,n_i\,v_s^2\ ,
\end{equation}
in agreement with the velocity observed in our simulation \cite{Kemp12, Ping12}. As the plasma flows inward it can filament \cite{Ren04}. Note that in the limit of total laser absorption the approximations made here fail: the electron rest mass, as well as ion energy flux are not negligible. Typical absorption fractions observed in intense laser plasma experiments are $f_a<0.9$, which is an upper limit because it includes re-absorption of the reflected light in under-dense plasma.

The total laser absorption fraction and the low-energy part of laser-generated electron distribution function (EDF) remain relatively constant for several picoseconds. Figure \ref{fig:spectrum} shows the electron energy flux along the laser direction for all (black curve) and for $E<1.5E_p$ (blue curve) electrons, and the net laser flux through the box boundary; all quantities are normalized to peak laser power, which would amount to 1.3PW when rotated around the symmetry axis. The difference between total electron energy flux and net laser flux is due to the projection of the electron velocity on the horizontal axis. The red curve shows an increasing amount of electron energy flux in an energetic population that is formed due to the stochastic acceleration of electrons in the expanding plasma in front of the target.

\subsection*{Stochastic acceleration of electrons in a large-scale density gradient}
\label{sec:stochastic}

Situations where the relativistic laser pulse interacts with large volumes of sub-critical density plasma can result either from an energetic pre-pulse that contains a small but finite fraction of the energy contained in the main pulse \cite{MacPhee10}, see Sec. \ref{sec:preplasma}, or from the early phase of a multi-picosecond interaction of the main pulse itself. The PIC simulation shown in Fig. \ref{fig:spectrum} \cite{Kemp12} corresponding to the latter scenario shows that the electron density profile resembles an isothermal expansion $n_e(z,t)=n_{e,0}\exp{-z/c_st}$ with $n_{e,0} =0.15n_c\,a_0^{1/2}$, where $n_{e,0}$ is determined by details of the surface emission and a sound velocity 
$c_s=\sqrt{(m_ec^2/M_i)a_0}\simeq0.05c$. 
The power $P_{ex}=8m_p n_{e,0} c_s^3$ driving such a self-similar expansion \cite{Atzeni04} amounts only to small fraction of the incident laser power, but the presence sub-critical density plasma increases the electron population accelerated through stochastic heating to energies of tens of MeV. 

While the equation of motion of an electron in a single laser pulse is regular and can be solved analytically \cite{Sarachik70,Quesnel98}, its motion in two counter-propagating pulses can become chaotic if the amplitudes are sufficiently high \cite{Mendonca83, Forslund85, Sheng02}. This means that individual particles can gain energy beyond the ponderomotive potential and a quasi-thermal distribution evolves. Recent work related to this so-called stochastic heating and acceleration (SHA) effect \cite{Sheng02, Bourdier07, Kemp09} has demonstrated that plasma at a few-percent of critical density facilitates this mechanism by providing plasma waves or an electrostatic potential well. This is because electron scattering off the quasi-static electric field enhances the stochasticity of its motion. In addition, Raman scattering in plasma provides backscattered light that acts as a counter-propagating pulse even when there is not an external secondary light source \cite{Forslund85, Kemp09}. 

\begin{figure}[t!]
\begin{center}
\includegraphics[width=.5\textwidth]{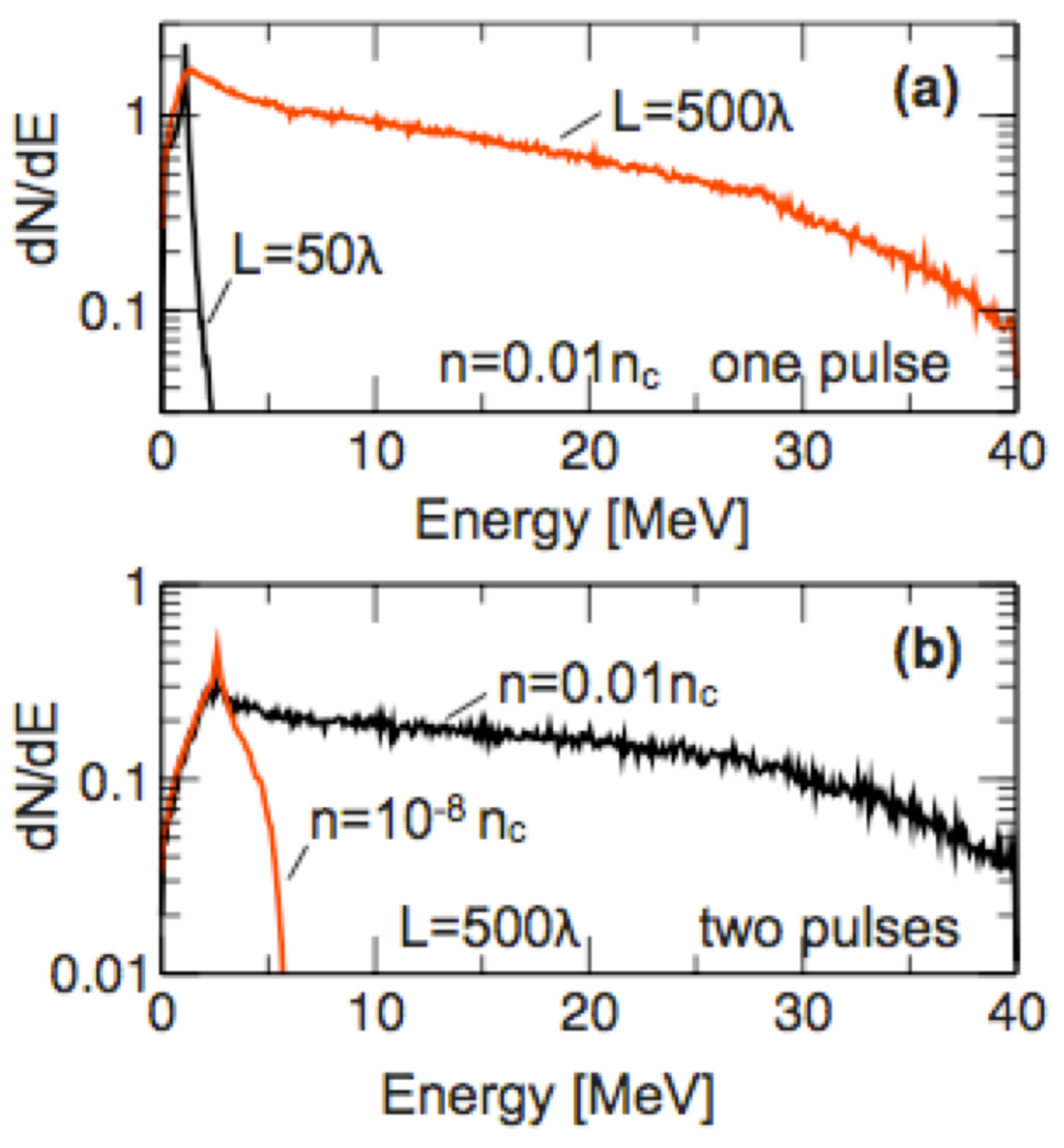}
\caption{\label{fig:stochastic} Role of longitudinal electric field for underdense plasma interaction. Electron spectra for different plasma length $L$, density $n$, and one/two pulse(s) at $I\lambda_L^2 = 10^{19}$ W/cm$^2$; (a) one pulse, vary plasma length; (b) two pulses, vary plasma density.}
\end{center}
\end{figure}

Fig. \ref{fig:stochastic} demonstrates the effect of stochastic heating and acceleration (SHA) under different plasma density and -length conditions in 1D PIC simulations. In Fig.\ref{fig:stochastic} (a), one pulse (at 1 $\mu$m wavelength, 10$^{19}$ W/cm$^2$) is injected into a plasma at 1\% of the critical density, while the plasma length is varied between 50 and 500 $\mu$m. The electron spectrum becomes 'hotter' with increasing plasma length because Raman backscatter generates a counter-propagating pulse, which facilitates SHA over increasing distances. In Fig. \ref{fig:stochastic} (b), two pulses are injected in from opposite directions into a 500 $\mu$m long plasma, while the plasma density is varied. For a vanishing plasma density, i.e., at $n_e=10^{-8}n_c$, the electron currents are negligible compared to the laser field. Here electrostatic effects are effectively suppressed so that the spectrum is limited to the ponderomotive energy. At a density of 1\% of the critical density, the electron spectrum for the case of two pulses resembles the case with only one laser pulse shown in Fig  \ref{fig:stochastic} (a). In order to prevent Raman backscatter of a single pulse generating a secondary pulse in our one-dimensional plasma model, modified simulations were performed in which the transverse plasma current is set to zero at each time step. For a single injected laser pulse, these modified simulations give electron spectra that resemble the case with vanishing plasma density shown in Fig.  \ref{fig:stochastic} (b). When two counter-propagating pulses are injected, however, we get a spectrum that resembles the one shown for finite plasma density in Fig. \ref{fig:stochastic} (b).  

Competing with SHA as acceleration mechanism in under-dense plasma are (1) Raman Forward Scattering \cite{Joshi81, Forslund85}, where the laser drives a strong plasma wave that moves in phase with the laser pulse and accelerates particles to high energies (2) direct laser acceleration (DLA) \cite{Pukhov99,Naseri12}, where electrons scatter off electromagnetic field structures on the side walls of the laser formed plasma channel, and (3) resonant absorption \cite{Freidberg72}. We find that these mechanisms play an insignificant role compared to SHA in situations relevant to fast ignition. LWFA occurs ideally for few-cycle pulses travelling through uniform plasma with well-matched laser and plasma parameters. The comparatively long pulse durations and large plasma density gradients found in fast-ignition relevant cases lead to an overwhelmingly stochastic, non-coherent acceleration process in fast ignition scenarios. DLA, on the other hand, depends on the interplay between the laser electric field and electrostatic fields at the walls of the laser-created channel in sub-critical density plasma. To investigate the role of DLA in a situation typical for FI, we have performed 2D kinetic PIC simulations of a fast-ignition relevant laser pulse interacting with a $20 \mu$m long and $20 \mu$m wide uniform shelf of under-dense plasma followed by a region of over-dense plasma with an absorbing boundary at the end. We compare two simulations, one with p-polarization, i.e., the laser electric field lies within the simulation plane, and the other one with s-polarization, where the laser electric field points out of the simulation plane. We find that electron energy spectra are almost identical at energies above the ponderomotive energy, $E_p$. This suggests that DLA does not play a significant role for electron acceleration, since it depends on the electrons interacting with the channel walls so that that a change of laser polarization would affect the amount of energy distribution, also see Ref. \cite{Cai10}. For SHA, on the other hand, laser polarization makes no difference on the spectrum, in agreement with our observation. We want to point out that in Ref. \cite{Kemp12} the acceleration of electrons in under-dense plasma should have been associated with SHA for the reasons given here. On the other hand, in 2D simulations the p-polarized case gives more electron energy flux at energies below $E_p$ than s-polarized irradiation. This is related to the fact that under s-polarized irradiation in 2D geometry the laser-driven electron currents are perpendicular to the simulation plane; they do not lead to charge separation and hence cannot cause an interplay between the laser magnetic field and fields caused by the electron beam filamentation near the critical interface. Resonant absorption (RA) should, similar to DLA, depend on laser polarization in 2D simulations, i.e., under s-polarization conditions it should be significantly reduced. Since our test simulation gives no significant difference in the electron spectra, we conclude that RA plays no active role in intense short-pulse laser interaction for fast ignition. 

The directionality of the laser generated electron beam is a signature of the acceleration mechanism and depends on the scale-length of the pre-plasma. Since stochastic acceleration in under-critical density plasma takes place over several laser wavelengths, the resulting electron beam follows the laser irradiation direction; on the other hand, acceleration near the plasma surface occurs over less than one laser wavelength, so that the mean direction of laser-generated electrons is dominated by the surface normal. This is consistent with experimental observations by Santala et al \cite{Santala00}. They performed experiments at RAL's Vulcan laser system with 20-50J of energy delivered over 1ps, with p-polarized geometry under a 45$^\circ$ angle of incidence on solid targets, and found that the gamma-ray beam generated by the fast electrons moves from the target normal to the direction of the laser irradiation as the scale length of the pre-plasma in increased. However, recent experiments on LLNL's Titan laser by Chen et al. \cite{Chen13} with 150J of energy delivered over 0.7ps, i.e. at a ten times higher intensity than Santala's experiment, show an additional trend. Chen et al's experimental results indicate that magnetic fields generated by the laser interaction in under-critical density plasma can scatter a significant portion of laser-accelerated multi-MeV electrons away from the direction of the laser. This phenomenon is currently under investigation \cite{Chen13, Perez13}. 

Many published PIC simulations of intense short pulse laser interaction feature quasi-static magnetic fields in the expanding plasma surrounding the laser spot. Since these fields have strength on the order of the laser magnetic field itself, they can potentially affect the trajectories of MeV electrons. These fields can potentially play an important role for the interpretation of current short-pulse experiments \cite{Mondal12}, and several authors claim that they will affect the divergence of the electron beam in cone-guided fast ignition \cite{Sentoku04, Nakamura07, Micheau10}. In PIC simulations of large-diameter laser pulses with slab targets, however, magnetic fields play only a minor role due to the fact that they are mostly present at the edge of the pulse. 

Integrated simulations of cone-in-shell targets \cite{Chrisman08} suggest that at high laser intensities a large fraction of the absorbed laser energy would go into a sub-ponderomotive electron component that is generated with a density corresponding to the relativistic critical density and a reduced temperature. The characteristic energy of this population is \cite{Chrisman08}
\begin{equation}
\epsilon_h=m_ec^2(\gamma_{os}-1)\sqrt{\gamma_{os} {n_c/n_p}},
\end{equation}
where $\gamma_{os}=\sqrt{1+a_0^2}$, which is equivalent to the so-called $J \times B$ acceleration scaling at the relativistic critical density \cite{Wilks97}. Although this would be a useful feature for fast ignition to adjust the electron energy by changing the target density, recent simulations with higher intensities and longer pulse interactions shows that the sub-ponderomotive electrons disappear because of the surface deformation and the strong magnetic fields, which enhance the absorption, at later time. 

From the picture described above emerges the notion that the total absorption and the electron spectrum scales mainly with the normalized laser amplitude $a_0$. However, the ratio between the laser-spot size and -wavelength, as well as pulse duration to laser period and ion- to electron mass affect the relative importance of the various absorption modes with respect to each other.

\section{Characterization of the electron source}
\label{sec:source}

\begin{figure}[t!]
\begin{center}
\includegraphics[width=1.\textwidth]{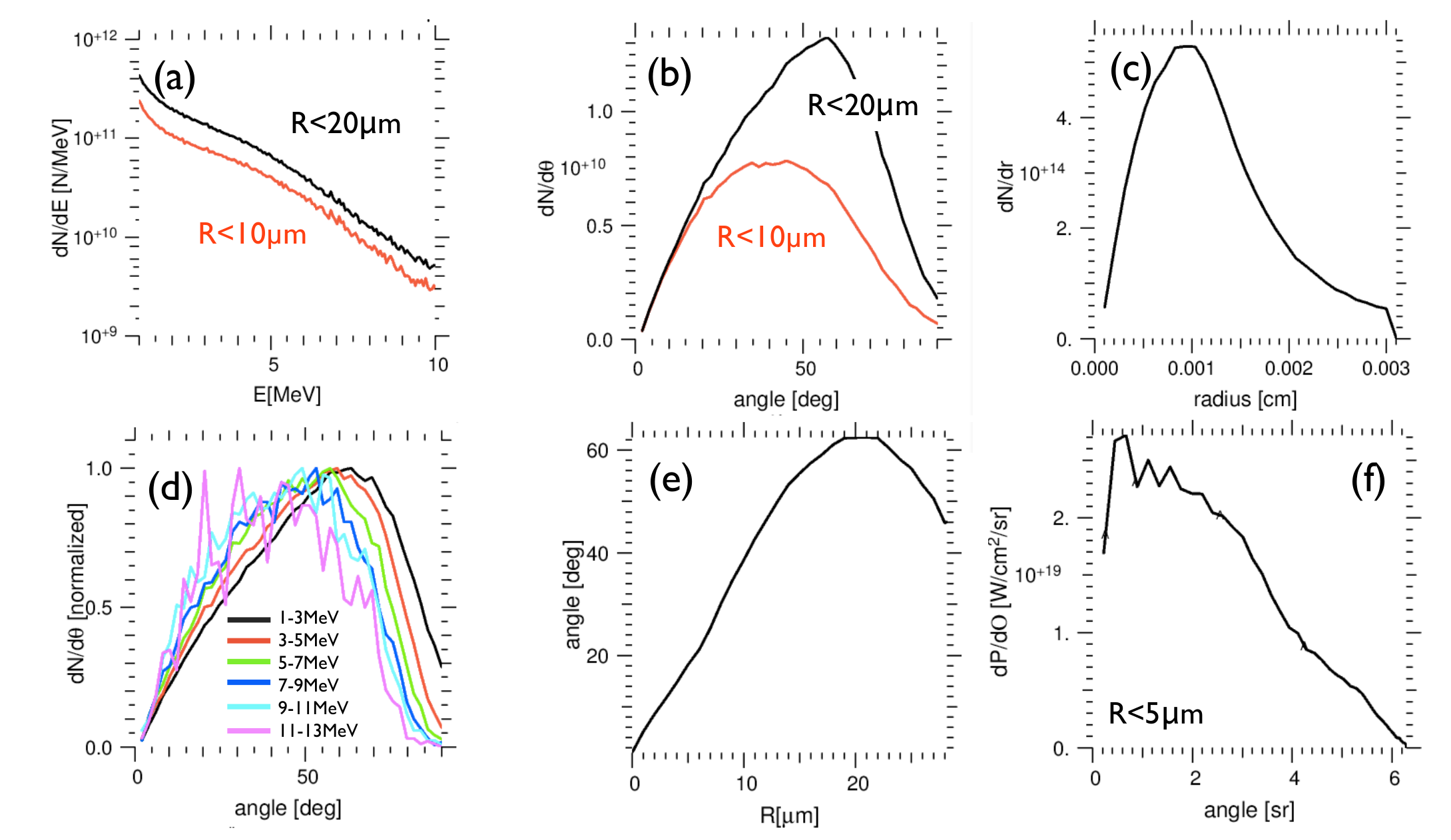}
\caption{\label{fig:source} Characteristics of the MeV-electron source measured in a 3D simulation in a disk located at z = 40 $\mu$m at t = 1200fs, under similar conditions to those shown in Fig. \ref{fig:spectrum}. Shown are (a) electron number spectra, averaged over two different radii; (b) angular distribution in particle number per 2$^\circ$ intervals; (c) particle number in $1\mu$m rings; (d) energy dependence of angular distribution; (e) mean angle vs radial position; (f) brightness averaged over a $5\mu$m disk.}
\end{center}
\end{figure}

In order to characterize properties of the electrons source quoted in Sec. \ref{sec:surface} in more detail, we have performed three-dimensional simulations similar to the 2D case discussed in Sec. \ref{sec:surface} \cite{Kemp12}. For economic reasons the laser spot was scaled by one-half in diameter, while its transverse and temporal profile, as well as the plasma density are identical to the 2D case. We have further reduced the box size to $40\times40\times60 \mu$m$^3$, leaving only 30 $\mu$m of vacuum in front of the target. The numerical resolution in the 3D run has been reduced to 16 cells per micron and it uses 15 particles per cell, maintaining numerical stability with third-order shaped particles and current smoothing. Comparable 2D simulations with the same parameters agree with this, confirming the viability of our approach. Consistent with the 2D case presented in Fig. \ref{fig:spectrum} above, the 3D simulation gives a coupling efficiency of about 25\% from the incident laser intensity into a forward going electron energy flux of particles $<$7MeV. Figure \ref{fig:source} presents a characterization of the electron distribution function (EDF) in a cylindrical disk with radius 20 $\mu$m and thickness 1$\mu$m, located 10 $\mu$m behind the original interaction interface (this location is chosen to avoid the region in which ions are accelerated into the bulk plasma and a co-propagating electron 'cloud') \cite{Denavit92, Sentoku03, Silva04, Fiuza12a, Fiuza12b}. While the energy flux at particle energies $<$ 1MeV is insignificant, i.e. it amounts to less than 10\% of the total electron energy flux at that time, future work will address details of the spectrum with better accuracy than possible today, which might have relevance for maintaining a realistic spectrum in the fast-ignition relevant energy window 1-3MeV after scaling the distribution to higher ponderomotive energies. In terms of coupling efficiency, overall divergence and energy spectrum between 500 keV and 20 MeV our results are similar to those presented by Debayle et al. \cite{Debayle10a}. We note that the use of wide laser pulses, with a spot radius of 20 $\mu$m or greater, as in these simulations, is important for the stability of the laser-plasma interaction and for the control of the electron beam divergence: 3D simulations performed with narrower laser pulses (5 $\mu$m spot radius) show a strong channeling and self-focusing in the overdense plasma, causing oscillations in the directionality of the fast electrons and a larger divergence \cite{Fiuza11}.

\begin{figure}[t!]
\begin{center}
\includegraphics[width=1.\textwidth]{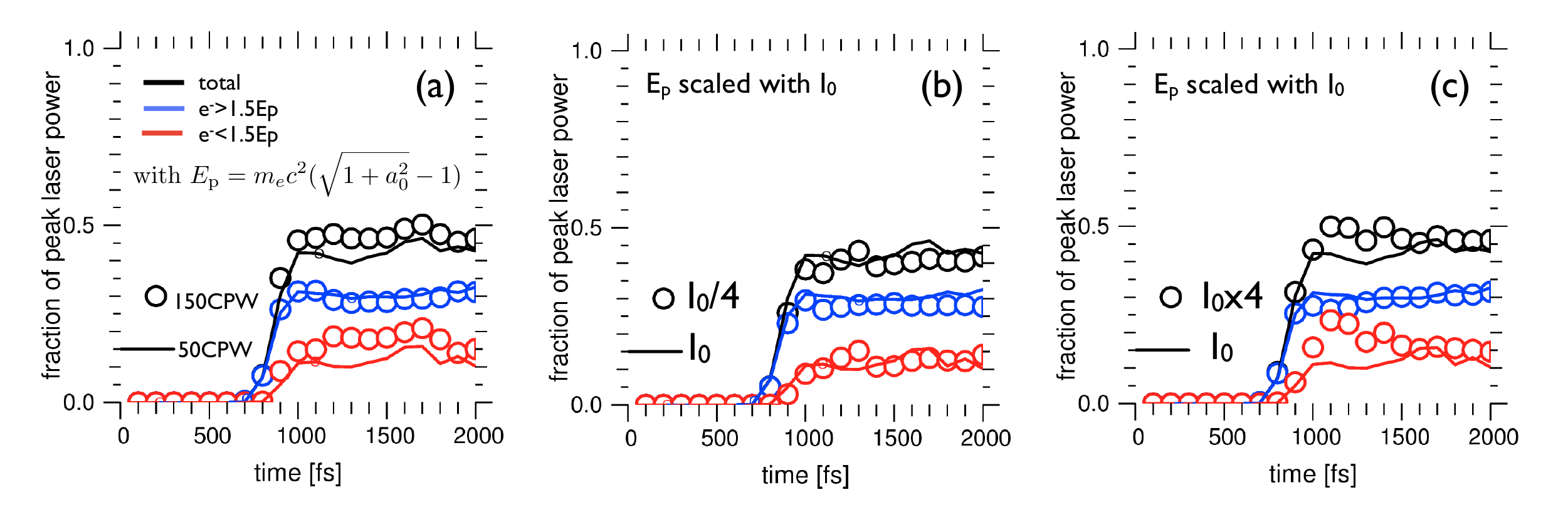}
\caption{\label{fig:scaling} Numerical convergence and scaling with intensity, in terms of electron energy flux density, measured in a plane behind the absorption layer. Shown are three energy groups: total (black); for electrons with kinetic energy below $1.5\times E_p$ (blue); and above (red). (a) Comparison of two runs at the nominal resolution (50 cells per wavelength) and at three times higher resolution; (b,c) scaling of the central result at intensity $I_0 = 1.4\times10^{20}$W/cm$^2$ at $1\mu$m wavelength (solid line) with intensity. Energy groups are scaled with respect to the ponderomotive energy $E_p$.}
\end{center}
\end{figure}

Numerical convergence of our 2D simulation shown in Fig. \ref{fig:scaling} has been verified with simulations at a spatial resolution of up to 150 cells per wavelength and 30 particles per cell in a $24\times 75 \mu$m$^2$ simulation box with periodic boundary conditions. We found that a simulation box width of less than ten laser wavelengths will lead to an underestimation of the surface emission effect because of the limited number of spatial surface modes, compare Fig. \ref{fig:scaling}. 

Scaling the results shown in Fig. \ref{fig:scaling} to higher or lower laser intensities or shorter laser wavelengths can, at least in principle, be done using the ponderomotive energy $E_p$, as shown in Fig. \ref{fig:scaling} for the case of intensity scaling. These simulations use a $24 \times 75 \mu$m size simulation box with periodic boundary conditions. The agreement between the central case at $I_0=1.4\times 10^{20}$W/cm$^2$ and the cases with $4\times$ higher or $4\times$ lower intensity demonstrates the robustness of the surface emission phenomenon discussed in Sec. \ref{sec:acc}. In corresponding full-scale simulations we expect that changing the laser intensity will affect the rate at which the vacuum region in front of the target is filled with plasma and thus the rate at which higher-energy electrons are accelerated through stochastic acceleration. However, the simple scaling with the ponderomotive scaling might break when new physics, eg. the formation of collision-less shocks behind the interaction region, comes into play. At several times higher laser intensities of around $8\times10^{20}$W/cm$^2$, such shocks have been demonstrated in simulations \cite{Fiuza12a,Tonge09} and their role on the fast electron distribution in fast ignition conditions needs to be carefully addressed.

When this characterization is used for electron transport simulations it is important to include the detailed radial profile, as opposed to using spatially averaged distributions with a uniform angular divergence.  As pointed out by Debayle et al \cite{Debayle10a}, the mean angle of the local EDF increases with radial distance from the beam axis, compare Fig. \ref{fig:source}. This angle can be explained as the viewing angle of a finite-size electron source seen from an observer plane at a short distance. Debayle et al \cite{Debayle10a} have demonstrated that ignoring this radial dependence of the mean angle can lead to an over-estimation of magnetic self-collimation of the laser-generated electron beam in transport simulations. In PIC simulations of an intense laser pulse at an intensity of $2\times10^{20}$ W/cm$^2$ and a Gaussian profile with a full-width half maximum of $20\mu$m interacting with a cone target, they quote a coupling efficiency of 35\% into electrons at energies $>$ 200keV, and an overall Gaussian $1/e$ beam divergence of 55$^\circ$, similar to the result quoted above. The overall divergence is obtained by integrating the product of radial profiles of mean angle, angular spread and beam density \cite{Debayle10a}. The electron energy spectrum is fitted by a power-law and resembles the EDF published by Kemp et al \cite{Kemp12}, shown in Fig. \ref{fig:source}.

The angular beam distribution quoted above is slightly different from the usual metric applied in experiments, and is important to relate them. Measured at observation planes with an increasing distance to the electron source, what is the opening angle of a cone that consists of the points where the beam intensity is half its peak value? We call the pitch angle of the cone the electron beam divergence. The most idealistic case is that of an isotropic point source emitting electrons into the forward direction, which has, when measured in the detection plane, a beam divergence of 45$^\circ$. This angle results from the drop in beam intensity with distance $I \sim 1/R^2$ so that $I_{plane}(r)=I_0 \times \cos^2(\theta)$ where $\theta$ is the angle between a line connecting the source with a point in the observation plane and the surface normal vector of the plane. If the source had an angular characteristics like the one shown above $P(\theta)=P_0 \exp[-(\theta/\theta_0)^2]$ with a $1/e$ angle of $\theta_0=57^\circ$, the corresponding beam divergence is 32$^\circ$, i.e. $I(0) \exp[-(\theta/\theta_0)^2] \cos^2(\theta)=I(0)/2$ for $\theta = 32^\circ$. 

On the other hand, for a point source to have a beam divergence of 20$\circ$, as suggested in experiments by Stephens et al \cite{Stephens03}, one would need a Gaussian angular distribution of the source with a $\theta_0=28^\circ$, smaller than observed in simulations. Note that in Stephens et al.'s experiment \cite{Stephens03} the distance between source and measurement plane is much larger than the source size, which justifies the approximation of a point source. The assumption that the angular distribution has a Gaussian shape is motivated by results in Sec. \ref{sec:absorption}. Modifying this assumption will affect the relationship between angular distribution and beam divergence in a non-trivial way. 

In addition to this geometric effect, the electron beam divergence inside a resistive medium can differ from ballistic transport. Recent PIC simulations that include realistic density as well as ionization effects predict the formation of collimating magnetic fields inside the target, which could alter the characteristics of the electron beam \cite{Sentoku11, Chawla13, Scott12}. Using pure transport simulations, other authors \cite{Honrubia06a} find that in order to reproduce a beam with an effective propagation angle of $\sim20^\circ$, as observed in experiments \cite{Green08, Stephens04}, they needed to assume an initial angular distribution with a half-angle of around 50$^\circ$. Recent PIC modeling by Scott et al \cite{Scott12} indicates that a fast electron beam associated with electron acceleration in under-dense plasma can generate a magnetic field within the target that is strong enough to partially collimate the subsequent, more divergent beam of lower-energy electrons. 

\subsection{Effects of pre-plasma}
\label{sec:preplasma}

The dynamics of the laser-plasma interaction depends on the pre-plasma profile that the high-power laser interacts with. This profile is determined by the amount of energy that leaks out of the laser's amplifier chain before the main pulse, as characterized by the laser system's energy contrast. Today's Petawatt systems deliver 1kJ of energy with an energy contrast of $\approx 10^{-7}$ and an intensity contrast of $\approx 10^{-10}$ \cite{LLE}. Contrast due to amplified superfluorescence and spontaneous emission (ASE) is independent of the final energy in the laser pulse. Therefore, for the ignition pulse of 100kJ mentioned above, the pre-pulse energy on target could range between 100mJ to 1J. Delivered over a nanosecond time scale at an intensity above $10^{11}$ W/cm$^2$ on target, this energy is sufficient to ionize matter and drive a plasma expansion into vacuum before the main pulse arrives. This leads to the formation of plasma with a multi-exponential density profile, determined by characteristics of the pre-pulse, the target geometry and -material. Details of these characteristics can only be determined in detailed hydrodynamic simulations, see the next Section.

In the last years several groups have performed PIC studies of the effect of pre-plasma on the laser absorption and particle acceleration \cite{Johzaki07, Baton08, Cai10, MacPhee10, Johzaki11, Johzaki12, Akli12, Sakagami12, Yabuuchi13, Li13, Ovchinnikov13}. It has been shown that an increase in the amount of pre-plasma not only places the absorption region further way from the core, leading to a reduced number of fast electrons reaching the core due to their divergence, but it also changes the dynamics of the laser propagation and the characteristics of the laser-generated fast electrons. 

As the ignition laser interacts with an extended pre-plasma, it can self-focus and filament, producing highly energetic and divergent electrons, which are not desirable for fast ignition of fusion targets \cite{Cai10}. As discussed in the previous Section, the laser stochastically accelerates electrons to very high energies in the extended plasma profile that forms in front of the target. Experiments with picosecond-scale laser pulses and corresponding PIC simulations have shown that a large scale-length pre-plasma will increase the interaction time of the laser with the under-dense plasma. This gives an enhanced number of electrons populating the energetic tail of the spectrum, and a decreased number of electrons in the energy range of interest for fast ignition due to pump-depletion of the laser pulse \cite{Baton08, Cai10, Johzaki11}. Cai et al \cite{Cai10} find that ASE-induced plasma extending 30-100$\mu$m in front of the target can reduce the forward-going energy flux of fast-ignition relevant electrons with energies $\leq 5$MeV to 10-50\% of its value with no pre-plasma, depending on the length of the pre-plasma. 

\subsection{Effects of cone geometry}
\label{sec:cone}

The idea of inserting a reentrant cone into the fuel shell was conceived in order to prevent potentially deleterious laser interaction with the coronal plasma and to minimize the distance between the interaction region and the compressed core \cite{Hatchett00, Kodama01a}. It avoids the difficulty of laser-driven hole boring into over-critical density plasma with another laser pulse, as envisioned in the initial fast ignition scheme \cite{Tabak94, Sentoku06, Li08}. In addition to maintaining a corridor close to the compressed core relatively free of plasma during the implosion, recent simulation studies have suggested a concentration of laser energy at the cone tip due to reflection of the laser beam off the cone walls, and enhanced coupling into fast electrons due to (a) transport of energetic electrons along the cone wall \cite{Sentoku04}; and (b) the provision of surface area at an angle with respect to the laser direction of incidence \cite{Lasinski09}. The latter applies to the macroscopic geometry of the cone target, as well as to structured surfaces, i.e. surface perturbations where wavelength and amplitude of the perturbation are comparable to the laser wavelength. 

On a fundamental level, 2D PIC simulations by Lasinski et al \cite{Lasinski09} have demonstrated that cone shaped targets give systematically higher laser absorption fractions than comparable flat targets, and produce electrons of higher energies. This advantage persists for flat-top cones, even when pointing errors are included. In addition to the increased absorption fraction found in macroscopic cone geometry, Lasinski et al find that structured flat targets give higher absorption than blunt ones. Periodic divots of up to 6um depth can enhance the absorption by more than 50\% compared to equivalent targets with a flat surface \cite{Lasinski09}.  

Electron guiding along the surfaces of short-pulse irradiated solid targets has been first discussed in collision-less PIC simulations \cite{Sentoku04} and later observed experimentally \cite{Li06, Habara06} and other PIC simulations \cite{Nakamura04, Nakamura07}. Recently Micheau et al \cite{Micheau10} have performed 2D PIC simulations of 100fs pulses interacting with cone targets. They find that, if there is initially no preformed plasma, electron transport along the cone walls leads to enhanced coupling into the cone tip. However, the effectiveness of electron guiding along the cone walls appears to be sensitive with respect to the scale length of the plasma density at which the laser is absorbed. With a pre-plasma at a scale length of 1/4 $\mu$m, as measured along the surface normal, they find that the guiding virtually disappears \cite{Micheau10}. 

The loss of guiding due to the expansion of plasma inside the cone shows up more clearly in simulations of laser pulses that last for more than a picosecond, the time scale on which ion motion becomes noticeable \cite{Levy11, Cottrill10}. The same conclusion was reached in experiments and modeling by Baton et al \cite{Baton08} who find that the coupling efficiency of the intense laser pulse with the cone tip may be severely degraded by the ASE induced pre-plasma. In fact, even ASE free interaction conditions have not resulted into any enhanced coupling in presence of a cone-attached target \cite{Baton08}. Comparing angular distributions of laser-generated electrons in simulations of flat top-cone and slab geometries, Lasinski et al \cite{Lasinski09} find good agreement between the two geometries and conclude that 'magnetic field guiding along sloping cone surfaces is not a key player for these energetic particles. 

Preformed plasma inside the cone, e.g. generated by the ASE pre-pulse, causes pump depletion of the laser pulse and increases the distance between the absorption region and the cone tip. Distance to the cone tip gives a stronger dilution of the electron beam and reduced coupling efficiency. This has been observed quantitatively in Ma et al. \cite{Ma12}, who irradiated a cone-wire setup with the Titan laser and measured both the absolute time-integrated K$_\alpha$ emission from the wire, as well as its spatial shape along the wire. The total K$_\alpha$ emission from the wire is used as an indicator of the energy that potentially exits the cone tip and contributes to core heating in a fast ignition scenario. They found that injecting an external pre-pulse before the intense main pulse can lead to a significant reduction in coupling efficiency, while the electron spectrum becomes hotter. In earlier experiments performed on LLNL's Titan laser with a 150J, 0.6ps main pulse at 1um wavelength, MacPhee et al. have explored the effect of preformed plasma by comparing two laser shots with and without an external pre-pulse \cite{MacPhee10}. From hydrodynamic simulations they conclude that a 100mJ level of ASE pre-pulse energy leads to the formation of a significant pre-plasma inside the cone. Their 2D PIC simulations of the experiment show that this plasma leads to a break-up of the main pulse into multiple filaments far from best focus and the cone tip. In effect, all of the laser energy is diverted away from the cone tip and the forward-going component of 2-4MeV electrons is eliminated by pump depletion in low-density plasma. A similar conclusion was reached by Baton et al \cite{Baton08}, who compared the brightness of a copper K$_\alpha$ spot measured in slab target geometry to that in a cone-on-slab geometry. Hydrodynamic simulations of the ASE prepulse of LULI's laser system demonstrate that the cone geometry leads to significantly longer-scale preformed plasma. This difference disappears as the ASE prepulse is drastically reduced through a non-linear frequency upconversion. 

Johzaki et al \cite{Johzaki11} have studied the effect of pre-plasma on the coupling efficiency to the compressed core of a fast ignition target with a combination of 2D PIC simulation of intense laser interaction and 2D Fokker-Planck simulations of electron transport in dense plasma. Figure \ref{fig:preplasma} shows the changes in the laser-generated electron spectrum observed in the cone tip with different exponential scale lengths of pre-plasma. While the spectrum shifts to three times higher energies (determined by the slope at $E>10$MeV) with increasing plasma scale length, the coupling into the fast-ignition relevant energy group of $<10$MeV electrons drops from 40\% with 1$\mu$m pre-plasma to 10\% with 10$\mu$m pre-plasma. Based on the electron spectrum observed in their 2D PIC simulation, Johzaki et al have then applied Fokker-Planck simulations of electron transport to the dense core about 60$\mu$m away from the cone. They find that the heating rate in the dense core drops by more than a factor three due to the pre-plasma. The detrimental effect of pre-plasma on the coupling efficiency to the core also further amplifies the consequences of a lateral misalignment (pointing error) of the main pulse with respect to the center of the cone tip \cite{Johzaki12}.

\begin{figure}[t!]
\begin{center}
\includegraphics[width=1.\textwidth]{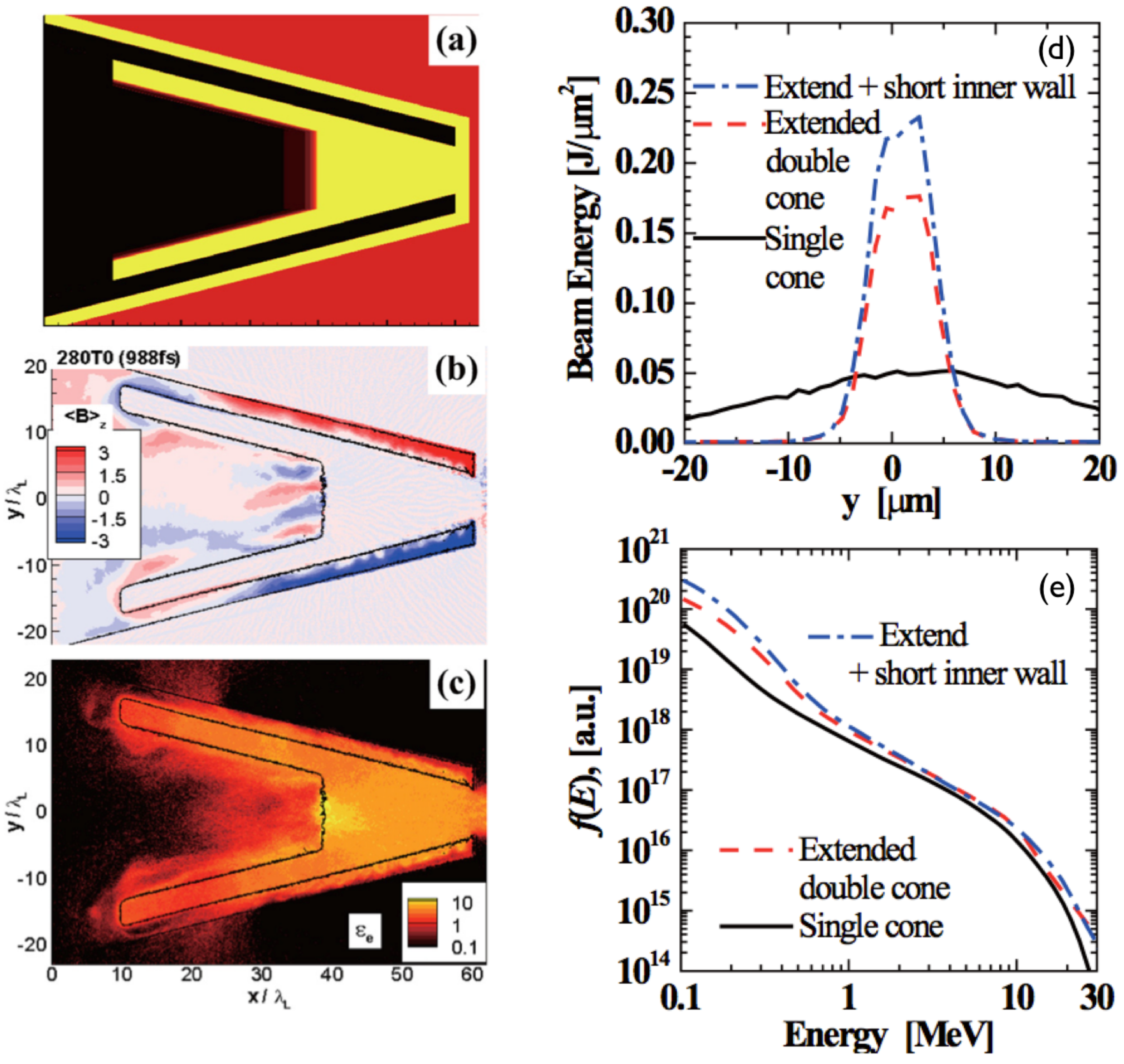}
\caption{\label{fig:cone} Double-wall cone targets give improved coupling efficiency compared to single-wall cones \cite{Johzaki11}. (left) (a) Initial density profile of the extended double cone with a short inner cone wall for 2D PIC simulations and spatial profiles of (b) quasi-static magnetic fields $<B_z>$ and (c) fast electron energy density $\epsilon_e$ at 280 $T_0$ (1ps). The lines in (a)-(c) show the density contours for $n_e = 10 n_c$. (right) (a) Transverse profile of time-integrated fast electron energy and (b) time- and space-integrated fast electron energy spectrum observed at $x = 62 \mu$m.}
\end{center}
\end{figure}

\begin{figure}[t!]
\begin{center}
\includegraphics[width=1.\textwidth]{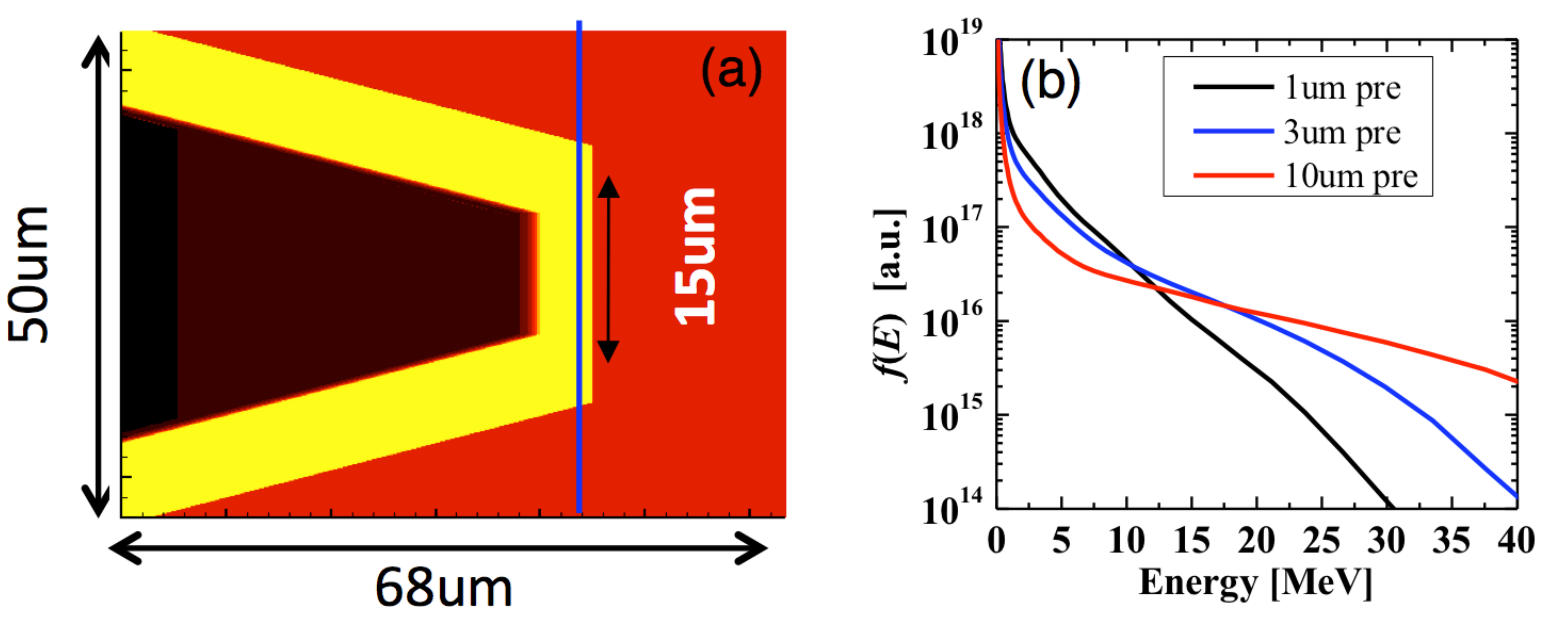}
\caption{\label{fig:preplasma} Effect of pre-plasma scale length inside cone target~\cite{Johzaki12}. Shown are energy spectra taken in the cone tip for three cases; (a) Initial electron density profiles. The cone plasma is assumed to be $Au^{40+}$ at $n_e = 100 n_c$ surrounded by $50 n_c$ CD plasma. The laser has a temporally flat and transversely Gaussian profile with $16 \mu$m FWHM at peak intensity of $3\times10^{19}$ W/cm$^2$ at 1 $\mu$m wavelength.}
\end{center}
\end{figure}

Extended double cones have recently been proposed to confine the fast electrons escaping from the cone by electrostatic and magnetic fields formed in the vacuum gap region of several micrometers width between the two walls. Johzaki et al \cite{Johzaki11} have demonstrated in combined PIC simulations of the laser interaction in the cone and Fokker-Planck simulations of electron transport that an extended double cone can enhance the core heating rate by more than a factor four compared to single cones, under otherwise equal conditions. Figure \ref{fig:cone} shows the geometry of double wall cone targets and the resulting magnetic field structure, as well as a comparison of energy flux and -spectra between a double-wall and a single-cone target. 

The obvious danger with the double-wall cone approach is that the hydrodynamic implosion prior to short pulse interaction destroys the double-wall structure. A further concern in the cone-guided approach to fast ignition is the use of high-Z material for the cone walls. While gold is the preferred material choice for a reentrant cone because of stability considerations during the capsule implosions, scattering of MeV energy electrons over tens of micrometers of gold inside the cone tip could lead to a loss in coupling efficiency due to an increased beam divergence or ranging-out of energetic particles. At the same time, hydrodynamic mixing of high-Z atoms with the core plasma could lead to intolerable energy losses through bremsstrahlung and line radiation. 

Extrapolating these results to full-scale FI configurations is difficult, however, because the detrimental effect of the expected pre-plasma could be mitigated by the hole boring associated with a more intense ($>10^{20}$ W/cm$^2$) and much longer ($\sim$10ps) heating pulse \cite{Baton08}. Assuming 80\% absorption, such a beam should be able to sweep away up tens of micrometers of highly ionized $10n_c$ Au plasma \cite{Baton08}; but this has not yet been demonstrated, neither experimentally, because of the enormous demands in terms of laser pulse energy, nor in simulations, because of the immense computational requirements, see Sec. \ref{sec:algorithms} - \ref{sec:experiments}.

\section{Advances in PIC algorithms}
\label{sec:algorithms}

The full-PIC modeling of the laser-plasma interaction in fast ignition relevant conditions is computationally very demanding. The spatial and temporal scales associated with the plasma oscillations must be resolved in the PIC code for accuracy, demanding $\Delta t \omega_{pe} \sim 1$ and $\Delta x \omega_{pe}/c \sim 1$ in the highest plasma density regions. In order to capture the interaction of intense lasers ($I = 5\times10^{19} - 10^{21}$ W/cm$^2$) with overcritical plasmas for time scales of the order of 1 ps, peak plasma densities of 100 $n_c$ are typically used, in order to guarantee that the plasma is opaque to the laser light, even when relativistic effects are taken into account, and that the laser light can only slowly push/hole bore the plasma. For this plasma density, the electron skin depth, $c/\omega_{pe}$, is 0.016 $\mu$m and the electron oscillation time, $1/\omega_{pe}$, is 0.05 fs. To resolve the skin depth with at least two points, the number of cells required to evaluate a 100 $\mu$m size plasma is 6250 in 1D, $3.9\times10^7$ in 2D, and $3.4\times10^{11}$ in 3D. For typical numbers of particles per cell (ranging from 1000 in 1D to 1 in 3D), this corresponds to advancing $10^7$ (1D), $5\times10^9$ (2D), 5$\times10^{11}$ (3D) particles for about  $4\times10^4$ time steps (1 ps), which leads to approximately $10^2$ (1D), $5.6\times10^4$ (2D), and $5.6\times10^6$ (3D) CPU hours. Even using high performance computing systems and highly optimized, massively parallel algorithms, multi-dimensional PIC simulations of the laser-plasma interaction for picosecond scales still require advanced numerical techniques.

In the last years, several numerical techniques have been developed and/or optimized for the PIC modeling of fast ignition (see e.g. \cite{Fonseca08} and \cite{Sentoku08}).

\subsection{Control of numerical heating}
\label{sec:heating}

A critical numerical issue in the PIC modeling of high-density plasmas for a large number of time steps ($\gtrsim10^5$) is grid heating, caused by under-resolving the plasma Debye length \cite{Birdsall}. Resolving the Debye length in all regions of the plasma can be extremely demanding from the computational point of view, in particular in high-energy density scenarios, where PIC simulations only resolve the collisionless skin depth, which is typically $10-100 \times \lambda_D$, and where grid heating cannot be neglected. The artificial heating of the plasma will significantly modify the properties of the background plasma, thus affecting the laser-plasma interaction and the transport of fast electrons.

The control of grid heating in the full-PIC algorithm can be achieved by using high-order interpolation schemes for the current deposition where macroparticles are represented by a cloud which extends for several grid cells and has an associated form factor \cite{Birdsall}. Additionally, current smoothing techniques can also be employed and help guaranteeing good energy conservation. A common technique for spatial filtering in finite difference PIC codes is of the digital type, where a given quantity $Q$ is calculated in cell $i$ using the value of cell $i$ and of the value of the adjacent cells $Q_iÕ = (W_1 Q_{i-1} + W_2 Q_i + W_3 Q_{i+1}) / (W_1 + W_2 + W_3)$, where the different $W$ represent the weight of each cell. Typically a binomial filter is used ($W_1 = 1, W_2 = 2, W_3 = 1$), which can be applied multiple times followed by a compensator. The compensator cancels the attenuation of order $O(k^2)$ near $k = 0$, allowing for a better energy conservation \cite{Birdsall}. 

\begin{figure}[t!]
\begin{center}
\includegraphics[width=1.\textwidth]{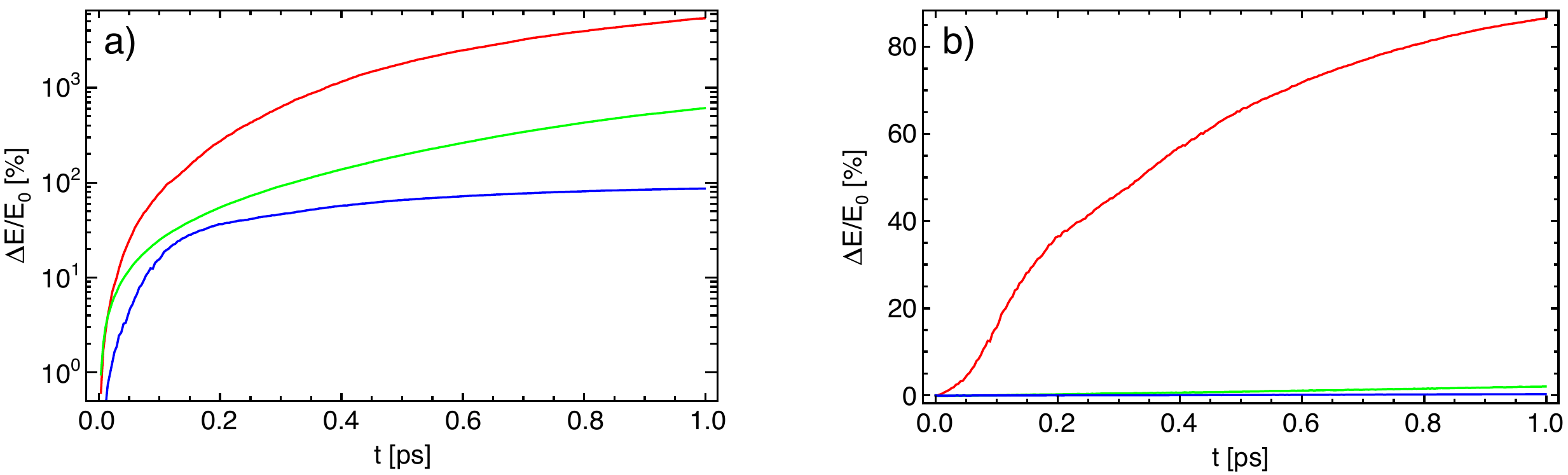}
\caption{\label{fig:numerics} Energy conservation in OSIRIS \cite{Fonseca02,Fonseca08} for typical fast ignition parameters as a function of the particle interpolation scheme. The initial plasma density is 100 $n_c$ and the initial temperature is 1 keV. a) After 1 ps, numerical heating leads to a variation of 5400\% of the energy in the simulation box with respect to the initial energy $E_0$ for $\Delta = 1.5 c/\omega_p$ and 16 ppc (red), 600\% with $\Delta = 0.5 c/\omega_p$ and 16 ppc (green), and 87\% with $\Delta = 1.5 c/\omega_p$ and 64 ppc (blue). b) Numerical heating can be dramatically improved using high-order splines. The increase of the energy in the simulation box using $\Delta = 1.5 c/\omega_p$ and 64 ppc is 87\% with linear (red), 2\% with quadratic (green), and 0.3 \% with cubic interpolation (blue).}
\end{center}
\end{figure}

Figure \ref{fig:numerics} shows the influence of the numerical parameters on the numerical heating for typical fast ignition parameters. We can observe that even resolving the electron skin depth and/or using a reasonably large number of particles per cell (64) is not enough to control numerical heating, but by using higher order particle shapes (in particular cubic and quartic interpolation orders) it is possible to guarantee that numerical heating is controlled to 1\% level for ps time scales \cite{Sentoku08, Fiuza11b}.  

\subsection{Anomalous macroparticle stopping}
\label{sec:stopping}

When modeling high-density plasmas, as the ones associated with fast ignition, it is common to use the technique of weighted macroparticles, meaning that each simulation particle has a charge density that can correspond to multiple real particles (electrons or ions). For instance, at a density of $10^{23}$ cm$^3$ (100$n_c$ for 1$\mu$m light), in a simulation with cell size of $0.5c/\omega_p$, and 16 particles (macroparticles) per cell, the density of macroparticles is $\sim 2.7 \times 10^{19}$ cm$^3$, and therefore each macroparticle represents $\sim 3700$ real particles. This technique enables the modeling of a given physical system with a reduced number of particles per cell, and therefore, in a more computationally efficient way. The majority of the plasma physics phenomena of interest depend on the q/m ratio, which is not modified by the use of weighted particles and thus the accuracy of the calculations is not affected by the use of this technique. However, there are a few physical mechanisms that depend on the exact charge of the particles, for which it is important to address the effect of using weighted particles. One such mechanism, of relevance for fast ignition, is the energy loss of charged particles due to plasmon emission (formation of wakefields in the plasma), which depends on $q^2/m$. Thus, it is important to address the influence of an increased energy transfer between fast electrons and the background plasma due to the large macroparticle weights in typical fast ignition simulations. 

In 2D slab geometry, a relativistic particle moving in a background plasma of density $n_p$, will lose energy at a rate of
\begin{equation}
\frac{d\epsilon}{dt} = -2\pi \omega_p q^2
\label{eq:stopping1}
\end{equation}
where $\epsilon$ is the energy of the test particle and $q$ its charge per unit length. We can observe that the energy loss depends directly on the charge and not just on the charge over mass ratio. In a PIC simulation, electrons have $q/m = 1$ and a charge that depends on the weight of the macroparticles, $q/e = n_p\Delta^2 S/N$, where $\Delta$ is the cell size, $N$ is the number of particles per cell, and $S$ takes into account the shape factor of the macroparticles. Defining $\Delta$ as the cell size normalized to the plasma skin depth, $c/\omega_p$, we can write the energy loss of a relativistic electron in a 2D PIC simulation as \cite{Tonge13}
\begin{equation}
\frac{d\gamma}{\omega_p d t} = -\frac{1}{4} \frac{\Delta^2}{N}S
\label{eq:stopping2}
\end{equation}
where $S \sim O(1)$ for $\Delta < 1$, i.e. when the plasma skin depth is correctly resolved. At solid densities ($\sim 100$ nc) the energy loss rate of a relativistic electron ($\gamma \gg 1$) plasma is $d \gamma/d\omega_p t \sim -4 \times 10^{-5}$. Thus, in a propagation distance of 50 $\mu$m its energy loss is negligible ($\Delta \gamma \sim 0.12$). However, in a PIC simulation this loss depends on the numerical parameters, and therefore in PIC simulations of fast ignition it is important to make sure that the cell size and number of particles per cell is chosen such that the energy loss of macroparticles in also negligible. The energy loss of a relativistic electron in 2D PIC simulations is $\sim 0.76 (\Delta^2/N)\sqrt{n_p/n_c}$ MeV/$\mu$m. In a typical 100 $n_c$, 50 $\mu$m plasma resolved with 2 points per skin depth, a large number of particles per cell ($N \gg 100$) must be used for the energy loss to be negligible. The use of high-order splines and current smoothing can, once again, help relax this constraint. 

We note that in order to control the energy loss of macroparticles, it would be useful to separate \emph{a priori} fast and background electron populations and use a higher number of particles per cell only for fast electrons. However, in the majority of the physical systems of interest, the evolution of fast and background electrons is dynamic and it is not possible to separate from the beginning of the simulation which electrons will be fast and which will belong to the background plasma.

\begin{figure}[t!]
\begin{center}
\includegraphics[width=1.\textwidth]{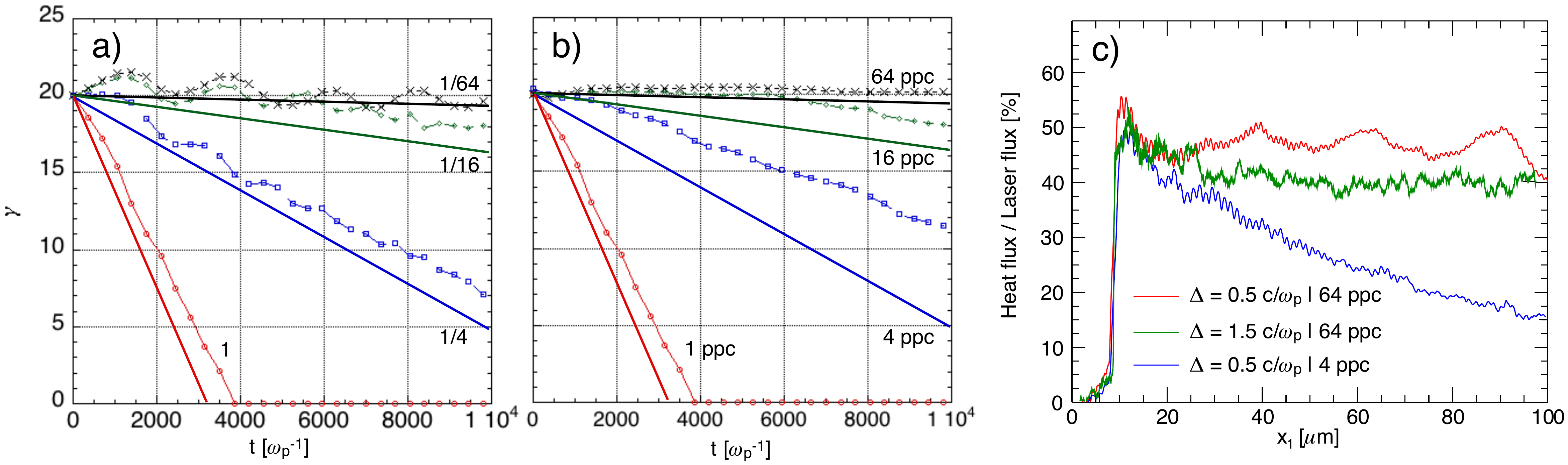}
\caption{\label{fig:macroparticles} Energy loss of relativistic macroparticles in overcritical plasmas. Comparison between OSIRIS PIC simulation results (markers) and the theoretical estimate of Eqs. \ref{eq:stopping1} and \ref{eq:stopping2} (solid lines) for the numerical energy loss of relativistic test electrons in a 100 $n_c$ plasma as a function of (a) the weight of test macroparticles ($n_p/N$) for fixed $N$ and (b) the number of particles per cell (ppc), N. (c) Fast electron heat flux for a fast ignition simulation, 1 ps after the interaction of a $2 \times 10^{20}$ W/cm$^2$ laser with a 100 $n_c$ plasma for different cell sizes ($\Delta$) and number of particles per cell (ppc).}
\end{center}
\end{figure}

Figure \ref{fig:macroparticles} illustrates the numerical energy loss both for a simulation where relativistic test electrons are launched in a 100 $n_c$ plasma and for a typical fast ignition simulation as a function of the numerical parameters. In can be seen that our theoretical estimate for the electron energy loss rate in plasma agrees reasonably well with the simulation results. For a small number of particles per cell the longitudinal electron heat flux (($\gamma-1)v_1/c$, where $v_1$ is the longitudinal velocity) is artificially reduced in the collisionless plasma. In order to have meaningful results the number of particles per cell cannot be smaller than 64 and the plasma skin depth must be resolved with at least 2 points \cite{Fiuza12c}.

\subsection{Boundary conditions and electron refluxing}
\label{sec:refluxing}

Another important numerical aspect of PIC simulation of fast ignition concerns electron reflexing and the appropriateness of the boundary conditions. In typical PIC simulations of fast ignition, the generated fast electrons propagate in plasmas with maximum densities of the order of 100$n_c$, for distances of 50-100 $\mu$m and are absorbed when they reach the simulation boundary. A critical aspect that needs to be understood is the effect of electron absorption at the boundaries of the simulation box. When a large number of fast electrons are either absorbed or thermally re-emitted from the boundary of the simulation box, large electric fields build up at the simulation boundary. This occurs for simulations times larger than the time it takes a relativistic electron to cross the simulation box, $L_{crossing} = L_{plasma}/c$, which is of the order of a few 100 fs, for typical plasma thicknesses $L_{plasma} = 50\mu$m. The large artificial electric field that is built up at the simulation boundary will start to reflect electrons back, causing the formation of a hot, relativistic return current that will modify both the transport of fast electrons and the laser-plasma interaction.

\begin{figure}[t!]
\begin{center}
\includegraphics[width=1.\textwidth]{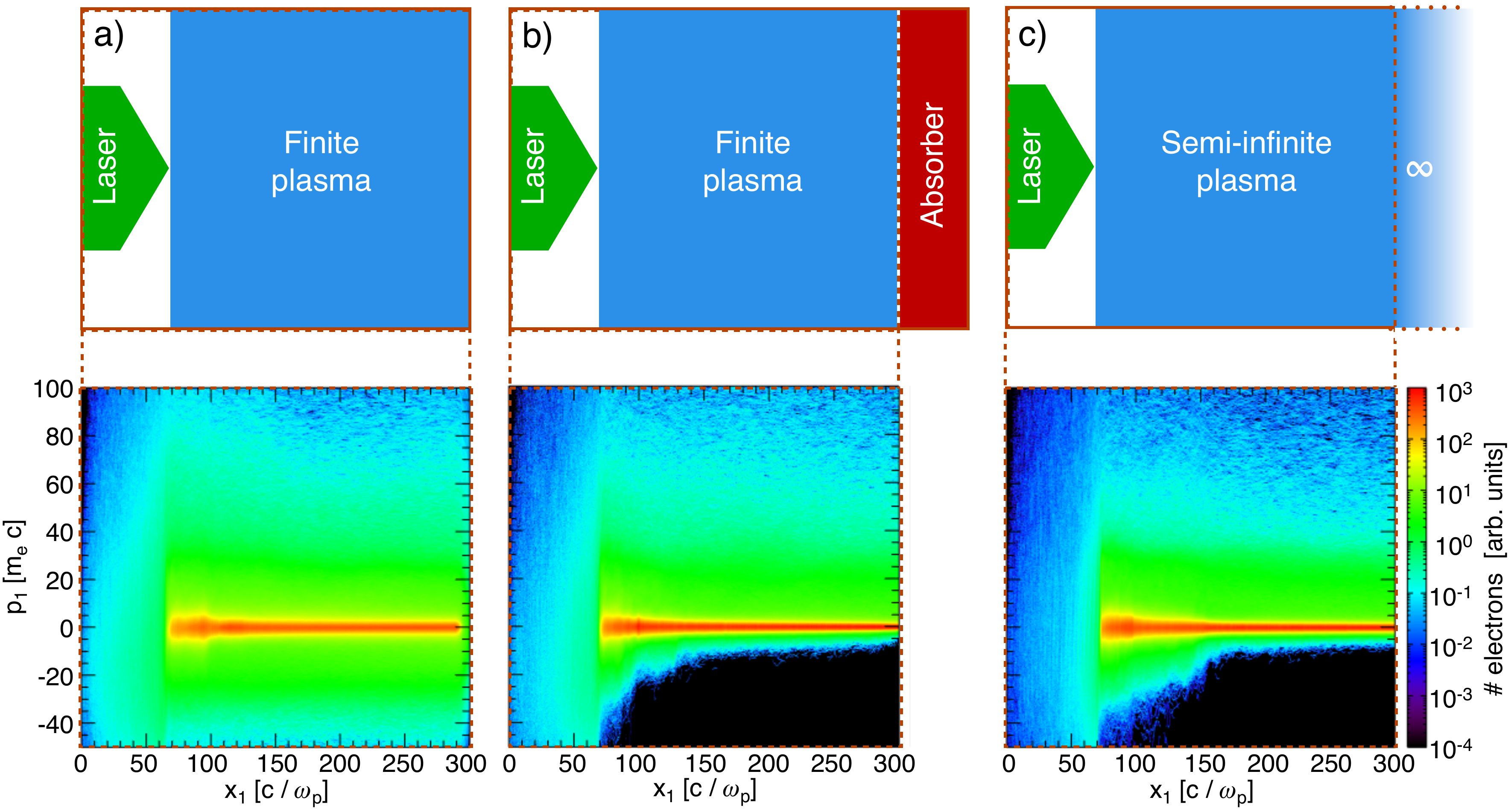}
\caption{\label{fig:absorber} Comparison of the return current properties in OSIRIS PIC simulations of the interaction of an intense laser with solid-density plasma for (a) a finite plasma with absorbing/thermal boundary conditions, (b) a finite plasma with an absorption region where particles are smoothly slowed down, and (c) a semi-infinite plasma. The use of an absorber prevents the generation of a strong electric field at the right boundary, avoiding refluxing, and leading to results consistent with a semi-infinite plasma setup for multiple picoseconds.}
\end{center}
\end{figure}

In order to perform full-PIC simulations of fast ignition scenarios for ps scales, it is crucial to avoid particle refluxing. A possible technique to avoid this refluxing is by having an absorption region before the simulation boundary, where fast electrons are smoothly slowed down, causing them to be absorbed without generating a large electric field \cite{Tonge09}. This absorbing region works as a special boundary condition where electrons suffer a drag proportional to their longitudinal momentum, that makes them stop or considerably slow down before they reach the boundary of the simulation box, where they are eventually absorbed \cite{Fiuza12c}. Figure \ref{fig:absorber} shows a typical electron phase-space for a full-PIC simulation of fast ignition where an intense laser ($I = 5 \times 10^{19}$ W/cm$^2$) hits a plasma with a density ramp from $n_c$ to 100$n_c$, followed by a 50$\mu$m flat region at 100$n_c$. We plot the phase-space for three different configurations: (a) standard absorbing boundary conditions at the end of the plasma, (b) an absorber region of 10 $\mu$m followed by the standard absorbing boundary conditions, and (c) a very long simulation box where electrons could not reach the end of the plasma (semi-infinite plasma). It is possible to observe that in the standard configuration there is a strong refluxing of electrons, which significantly modifies the return current. When the absorber is used, refluxing is avoided, and the return current remains cold, in very good agreement with the infinite plasma simulation. 

The modeling of realistic fast ignition conditions also requires isolated targets \cite{Tonge09} and that radiation is efficiently absorbed at the simulation boundaries, which can be obtained using perfectly matched layer (PML) boundary conditions \cite{Vay00, Vay02}. These techniques allow for the PIC modeling of fast ignition for multi-ps. 

\subsection{Coulomb collisions}
\label{sec:collisions}

In order to accurately model the transport of fast electrons in the high density plasma region (typically $> 100 n_c$) it is crucial to have an accurate description of Coulomb collisions between the different species of the plasma. Coulomb collisions are not captured in the standard full-PIC algorithm: it requires the development of a consistent collisional operator that allows for a correct description of the relevant statistical properties of the system. A common approach for introducing these effects in PIC codes is by doing binary collisions between macroparticles in a collisional grid using a Monte Carlo technique \cite{Takizuka77}. This method offers an exact solution of the Boltzmann equation and is ideal for the description of collisional effects in plasma  physics. However, the application of this method to high-energy density systems is not trivial, requiring advanced approaches to model in an accurate way the collisions between macroparticles with different weights (used to efficiently describe plasmas with strong density gradients) \cite{Nanbu98}, conserving both momentum and energy \cite{Sentoku08}, and being fully relativistic \cite{Peano09}. 

This type of collisional operator has been implemented and used in several PIC codes and used in the study of transport in fast ignition (see \cite{Perez12} for a detailed recent description of the implementation of this collisional operator in PIC codes)

\section{Towards multi-scale PIC modeling}
\label{sec:multi-scale}

The use of full-PIC codes to model the transport of fast electrons and the energy deposition in the high density region of a fast ignition target is outside the present capabilities, even with highly optimized algorithms and increasingly larger machines. The main limitation in explicit full-PIC codes is associated with the need to resolve the plasma oscillations in the entire simulation domain, $\Delta t \omega_p < 2$, for stability, and the Courant condition, $c \Delta t < \Delta x$. For typical core densities of a compressed fast ignition target, $\sim10^{26}$ cm$^{-3}$, this implies resolving temporal and spatial scales of attoseconds and Angstroms, respectively, and at the same time evaluate the dynamics of a mm size system for 10-20 ps. 

In the last years, there has been an increasing effort in developing advanced PIC algorithms to perform multi-scale modeling of fast ignition and couple the laser-plasma interaction with the transport and ignition. Different approaches have been followed with varying degrees of success.

\subsection{Coupling PIC with transport simulations}
\label{sec:transport}

The most common multi-scale approach is to use different algorithms to model different regions of the plasma. Full-PIC codes are used to model the laser-plasma interaction and to calculated the fast electron source for given laser parameters. This source is then used as an input for transport calculations with hybrid-PIC codes, that model the background plasma as a resistive MHD fluid and the fast electrons as kinetic particles \cite{Davies02, Gremillet02, Honrubia09}. This approach allows for the efficient coupling of the modeling at lower densities, where the laser-plasma interaction occurs, with the higher densities, where fast electrons are transported all the way to the core and resistively heat the background plasma. However, the kinetic effects associated with transport and the formation of return current in high-density plasma are not taken into account.

Computer models of electron transport in dense matter, e.g. ZUMA or similar models \cite{Larson09, Davies02, Gremillet02, Honrubia09} typically lack a description of the laser interaction, compare Sec. \ref{sec:absorption}; instead they require a prescription of an electron 'source' that is defined in a plane at one end of the simulation box. Electron sources can be provided by careful analysis of a PIC simulation of the laser plasma interaction, such as the one discussed in Sec. \ref{sec:source}. In the following we want to describe briefly how the transfer between the two codes can be carried out, making several assumptions about symmetry, temporal evolution of the interaction and its scalability with respect to laser intensity and wavelength. 

Transport simulations for fast ignition are typically performed in cylindrical geometry, i.e., they assume axial symmetry. This is justified by the observation that three-dimensional simulations of ignition-scale laser interaction show good symmetry with respect to the laser axis. They further assume that the source is in steady state, so that (a) microscopic fluctuations in the source behavior can be ignored and (b) there is no long-term evolution. Most current transport simulations are not using a temporal shape in the laser intensity. Modifying the laser intensity at a given laser spot shape for ignition scaling is done by scaling the characteristics extracted from PIC simulations, discussed in Sec. \ref{sec:source}, with respect to the ponderomotive energy \cite{Strozzi12}. Strozzi et al.'s \cite{Strozzi12} approach is to solve an inverse problem by 'guessing' an upstream source in a 'black box' and describing it analytically. Ballistic transport to the 'white box', located at the point where the distribution is 'measured' inside the PIC simulation, should then compare well to the 'measured' characterization. Bellei et al \cite{Bellei13} have suggested to use the characterization given in Fig. \ref{fig:source} above in connection with random sampling of particles as a source description. The characterization at a given time is manifested in a four-dimensional matrix that bins the electron distribution in the characterization volume vs. radial position; energy; pitch angle; and angle between radial vector and momentum. This 4D distribution is directly sampled via a Monte-Carlo technique for the injection from a plane in the transport simulation, which has the advantage that no fitting is required. 

We want to point out that smaller-scale diffraction-limited pulses typical for today's experiments show a different behavior where axial symmetry can be broken by quasi-static magnetic fields that form in the region where the laser is absorbed. These fields grow strong enough to scatter MeV electrons, and lead to hosing of the electron beam on 100fs time scale, which makes the interpretation of experiments more challenging \cite{Chen13}. 

The transition between the laser-plasma interaction and transport simulations require extra attention due to the mismatch between the two geometries at small and large radii. This mismatch is caused by (a) different focusing properties of a laser beam in two and three dimensions; (b) the power contained in the 'wings' of the pulse at a radius R and $R+\Delta R$ in plane 2D geometry is comparatively smaller than in a circular 3D spot because of the geometrical factor $2\pi R$. This means that it is not possible to simultaneously match the peak intensity and total power of a simulation in 2D cartesian geometry if the transverse coordinate is interpreted as a radial coordinate. Fast-ignition scale simulations presented here overcome this difficulty due to their relatively sharp radial drop in intensity $I \sim \exp{(-r/r_b)^8}$, which has relatively little power in the wings.  

\subsection{Full-PIC simulations with clamped density}
\label{sec:clampedt}

Sentoku and Kemp \cite{Sentoku08} suggested the possibility of modeling the different plasma regions associated with the fast ignition interaction with the same PIC structure, but where the plasma density is clamped to an artificial upper bound when computing charge and current densities in order to limit the plasma frequency and therefore the shortest scales to be resolved. For the purpose of computing collisions, the local electron and ion density are not limited, allowing for the description of collisional effects even at the core densities. However, the calculated electric fields associated with the plasma currents are not consistent with the actual density used for computing collisions and therefore the resistive heating of the plasma at high densities is inconsistent. This approach has been used by Chrisman et al \cite{Chrisman08} in PICLS to model the cone-in-shell ignition in 2D for up to 1ps. It has been shown that the core heating efficiency scales linearly with the laser intensity between $10^{19}$ W/cm$^2$ and $10^{20}$ W/cm$^2$. In these scale-down simulations, where the cone standoff distance is only 15 $\mu$m, the laser-core coupling efficiency is 15\% for $I = 10^{20}$ W/cm$^2$.

\subsection{Hybrid-PIC simulations retaining kinetic effects}
\label{sec:hybrid-pic}

More recently, the implementation of a hybrid algorithm in PIC codes has been suggested to allow for the modeling of the high density plasma regions while retaining kinetic effects, providing a consistent description of the different plasma regions \cite{Cohen10}. At low-density, high-temperature regions, close to the laser-plasma interaction region, where kinetic effects dominate, MaxwellÕs equations are solved as in standard PIC codes. At high-density, low-temperature regions, where collisional effects dominate, leading to strong damping of EM and plasma waves, an MHD system is used, coupling a reduced set of MaxwellÕs equations with a generalized OhmÕs law. By having both algorithms in the same PIC code structure all plasma species can be described with particles, not only the fast electrons, but also the resistive plasma. The fluid quantities required to advance the MHD system are calculated using the different fluid moments based on the self-consistent particle distribution function. This description allows for the generation of return currents with the correct distribution function, for the self-consistent separation between cold and fast plasma electrons, and for the accurate modeling of the energy exchange between different species (plasma resistive heating). The transition between full-PIC and MHD algorithms is done around a few 100 $n_c$, which is the density that determines the smallest scales to be resolved by the PIC code, allowing for great computational savings and for the modeling of realistic ignition scales. Kemp et al. have implemented this algorithm in PSC \cite{PSC} and used it to model the interaction of sub-ps laser pulses with cone-wire targets, of relevance for current experiments \cite{Kemp10}. Fiuza et al. have also implemented this hybrid algorithm in h-OSIRIS \cite{Fiuza11b} and have recently used it to model the interaction of a 100 kJ ignition laser with a compressed fast ignition target for the full density range and for realistic spatial (0.5 mm) and temporal (5 ps) scales, showing the possibility of reaching laser-core coupling efficiencies between 5-10\% \cite{Fiuza13}.

\section{Surrogacy of current experiments for full-scale fast ignition}
\label{sec:experiments}

The previous sections have described theoretical and simulation studies of laser-plasma interactions at ignition-scale, including the physics of intense light absorption and electron acceleration. Centrally important to fast ignition are the detailed properties of the electron distribution function, in terms of the overall flux, energy spectrum and angular distribution, predicted by theory or simulation.

A key aspect of advancing any theory or simulation and assessing its accuracy is experimental validation. While present-day experimental laser facilities can achieve the intensities of an ignition laser pulse, they can do so for only reduced-scale focal spots and pulse durations, because they are limited in the total delivered energy.  Furthermore, all experiments to date have been performed on laser systems designed to produce near-Gaussian intensity distributions in both space and time. Thus, the power spectrum incident on the target is not a delta-like function at a single intensity, but a broad distribution with significant power over a wide range of intensities, spanning more than an order of magnitude. While a few laser systems may have close to diffraction-limited performance, most will exhibit some level of aberration due to amplitude and phase non-uniformities in the incident beam or non-ideal focusing onto the target, resulting in a spatially aberrated beam in the focal plane further broadening the power spectrum in incident intensity.  It is important to not only validate the predictions of the interactions experimentally available but also assess the impact of these reduced spatial and temporal scales, and broad intensity distributions on the laser-plasma interaction process and resulting electron beam. 

The measurements most generally employed in experiments to study the interaction of intense light pulses with solid targets can be grouped into a few main categories: (i) measurements of the reflected laser light \cite{Ping08, Ping12, Palaniyappan12}, (ii) direct measurements of the fast electrons exiting the target \cite{Malka96, Yabuuchi07, Link11}, and (iii) indirect measurements of the fast electrons through their production of K$_\alpha$ radiation \cite{Stephens04,Perez10, Green08}, bremsstrahlung \cite{Hatchett00, Chen09},  coherent transition radiation (CTR) \cite{Storm08, Bellei12, Baton03}, or through target heating \cite{Lancaster07,Ma08,Nilson09}.

Measurements of the laser light reflected from the target surface, including the absolute reflectivity, shifting or broadening of the spectrum, harmonic generation, and changes in polarization, are sensitive to properties of the plasma below and up to the critical density surface and can provide information on the preplasma scale length, plasma motion, and magnetic fields at the absorption interface. They are therefore potentially very useful for validating LPI simulations. An example is Ping et al. \cite{Ping12} in which the authors performed a time-resolved measurement of the wavelength shift of the frequency-doubled light produced during the interaction of a $10^{20}$ W/cm$^2$ peak intensity, 700 fs pulse with a solid target. A 2D PIC simulation initialized with a laser focal spot distribution and preformed plasma based on experimental on-shot focal spot and laser contrast measurements reproduces well the magnitude and temporal behavior of the wavelength shift, relating it to the recession velocity of the critical density interface.
Direct measurements of the fast electrons exiting the target with electron spectrometers does not provide a direct measure of the initial electron distribution because the spectrum is modified by both transport in the target and more importantly time- and space-dependent electrostatic potential produced at the target surfaces due to charge separation \cite{Fill05}. A study by Link et al. \cite{Link11} showed that for typical laser energies of $\sim$100 J and target thickness of $\sim$10Õs $\mu$m the effects of the charge build-up and ion acceleration shifts the entire spectrum and weights the measured escaping spectrum to the early part of the pulse, while the surface potential is still building. While information on the low energy part of the original source spectrum is lost, the high-energy slope of the measured spectrum can be related to the original source spectrum after allowing for some change in the slope due to charging.

Indirect measurements, such as K$_\alpha$, bremsstrahlung, OTR, XUV and X-ray spectroscopy, require a transport model to relate the measurement to the fast electron distribution. Honrubia et al. \cite{Honrubia06} use a hybrid-PIC model with an analytic expression for the injected electron distribution to match K$_\alpha$ and rear-surface XUV measurements. Storm et al.  \cite{Storm09} use a similar approach to match 2D spatially resolved CTR. The challenge with this approach is that assumptions are required for the functional form of the injected electron distribution (for energy spectrum, angular distribution, and injected spatial profile). 
Chen et al. \cite{Chen09} use an alternative approach where the electron energy spectrum is not constrained by a predetermined model, but allowed to take any arbitrary two-temperature distribution. Several million Monte Carlo simulations are run with different electron energy distributions spanning a wide 3D space (two slope temperatures and the ratio). For each simulation the bremsstrahlung spectrum is calculated and a least-squares fit performed to a measured spectrum. Those electron distributions producing a reduced chi-squared fit value of less than one are deemed to be consistent with the experimental measurement. The conclusion, however, is that there is a large degeneracy in the injected electron spectra that yield the same experimentally measured bremsstrahlung spectrum. Ultimately, while such approaches can provide constraints on the electron distribution they cannot provide a unique distribution, independent of simplifying assumptions, which can quantatively validate the results of a PIC calculation.

In order to experimentally validate the accuracy of a PIC-calculated electron distribution one must ultimately proceed with a forward calculation where one models both the continued propagation of the electrons through the target and the measured observables. The difficulty in this case is the computational challenge of modeling both the laser-plasma interaction and electron transport in the solid target in a single simulation. Such simulations have been performed for thin targets ($< 20 \mu$m) and sub-picosecond pulses, enabling, for instance, escaping electron or proton spectra to be calculated \cite{Sentoku11}. However, in thin targets electron refluxing negates the notion of a forward-going electron beam with a well-described source distribution.

One method that has enabled LPI simulations with large targets at solid density makes use of the implicit-PIC scheme, as employed in the code LSP \cite{Welch06}. In this scheme the requirement for the Debye length to be resolved at high density is relaxed. Ovchinnikov et al.  \cite{Ovchinnikov11} have used the LSP code, in 2D Cartesian geometry, to study the interaction of an intense laser pulse with a 300 $\mu$m thick solid density target, including the calculation of K$_\alpha$ emission from buried layers in the target. They find that the spatial distribution of K$_\alpha$ does not directly correspond to that of the fast electron beam but is modified by electrons reflected from the target surfaces, including the front surface.

Accurate quantitative modeling of an experiment requires a 3D geometrical description of particle fluxes, currents, and fields. A full 3D explicit or implicit treatment of the interaction of a picosecond pulse with a large non-refluxing target at high density remains beyond current computational capabilities. A recent approach attempts to address this problem by simulating the laser-plasma interaction region with a 2D or 3D PIC calculation, sampling the electron distribution in a plane just beyond the absorption interface, and injecting this distribution as a source into a 2D axisymmetric or 3D hybrid-PIC calculation of the full target. This approach has been applied both in fast ignition design studies \cite{Strozzi12} and modeling of experiments \cite{Chen13}. Chen et al. have used it to quantatively compare the predictions of a 2D PIC calculation with absolute bremsstrahlung spectra recorded along three directions behind the target. The PIC simulation attempts to replicate the initial conditions of the experiment as closely as possible by initializing the laser phase profile such that the vacuum focal intensity distribution matches the on-shot experimentally measured intensity distribution, and initializing the preformed plasma with the output of a 2D radiation-hydrodynamics simulation of the measured laser prepulse. The choice of matching the incident intensity distribution means that there will be a small discrepancy in the simulated and measured spatial profiles because in a 2D Cartesian geometry it is not possible to simultaneously match both the spatial profile and intensity distribution. The electron distribution is recorded every 20 fs in a 1$\mu$m thick box a few $\mu$m behind the absorption surface, producing a 4D distribution (space, time, energy, angle). This distribution is mapped from 2D Cartesian to 2D RZ geometry and sampled to produce the electron source injected into a 2D RZ hybrid-PIC simulation using the Zuma code \cite{Larson09}. The Zuma simulation models the full spatial extent of the target, a 1.5 mm thick Al/Ag/CH multilayer with $5\times5$ mm lateral dimensions, and computes the K$_\alpha$ and directional bremsstrahlung emission produced. A problem arises, however, because the initial PIC simulation predicts a rather complex energy-dependent and time-varying directionality and divergence angle of the fast electrons. It predicts an electron distribution with two main components: (i) a broad, symmetric component centered on the target normal axis, and (ii) a narrower, asymmetric high-energy component with a time-varying directionality. This asymmetric component cannot be represented in 2D RZ geometry, which assumes axial symmetry. The authors incorporate the asymmetric component through a separate 3D Cartesian calculation made computationally tractable by running without self-generated fields. Comparison of the predicted and measured bremsstrahlung signals using a reduced $\chi^2$ test shows different levels of agreement for high and low energy parts of the spectrum. For electron energies $>2$ MeV the PIC-predicted absolute electron flux, spectrum, and angular distribution are all consistent with the data. For electron energies $<2$ MeV the data indicate a higher flux and larger divergence than predicted.

The value of the forward calculation approach is that it enables one to (i) establish if the PIC simulation is consistent with a given experimental measurement, within the measurement uncertainties, and (ii) determine the degree to which an experimental measurement constrains or validates the accuracy of the simulation. In terms of validating the properties of the fast electron distribution, for instance, one can perform transport simulations with variations in the flux, energy spectrum, and angular distribution and determine for each parameter the range that would be consistent with the experimental data.

While the 3D treatment of LPI and transport at full experimental scale is on the near-horizon, most modeling of laser-solid interaction experiments is presently confined to 2D. As we have seen this introduces a number of complications whose impact needs to be assessed, including the appropriate representation of a real focal spot in 2D Cartesian geometry, the validity of transferring an electron distribution between codes in the absence of feedback, the inability to treat a non-axisymmetric distribution in 2D RZ geometry.

\begin{figure}[t!]
\begin{center}
\includegraphics[width=1.\textwidth]{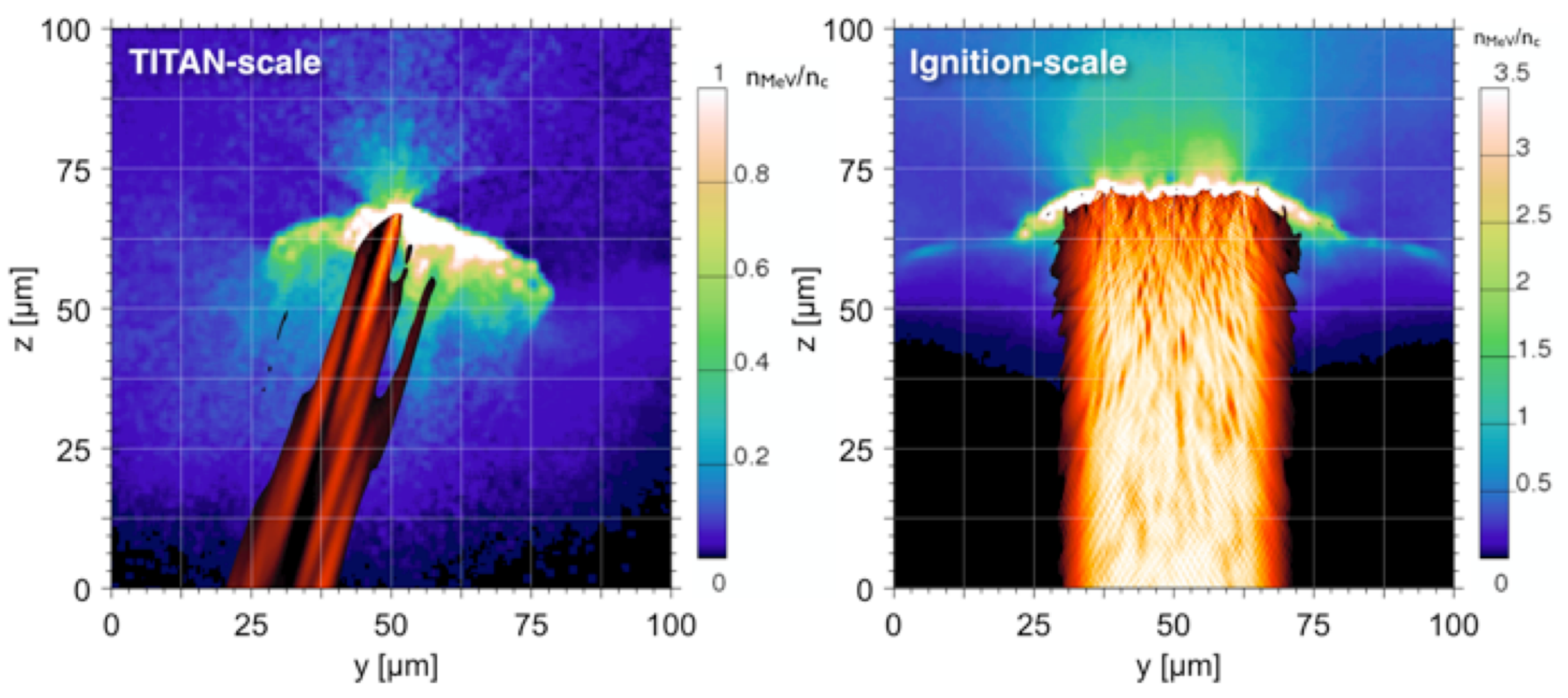}
\caption{\label{fig:surrogacy} (left) PIC simulations showing the interaction of the TITAN laser pulse, and (right) an ignition-scale laser pulse incident on a solid target; laser Poynting flux (red-black); energetic electron density (white-green-blue).}
\end{center}
\end{figure}

Given progress in the validation of LPI simulations with present-day experimental facilities one must assess the uncertainties in extrapolating simulations to ignition-scale. Figure \ref{fig:surrogacy} shows snapshots from 2D PIC simulations of the interaction of a laser pulse with a solid target using measured parameters of the Titan laser (left), and an ignition-scale laser (right). The extrapolation from one to the other involves consideration of spatial, temporal, and physics effects. The spatial or geometric effects are readily apparent - the beam phase distortions and f/3 focusing of the Titan pulse produce filamentation and self-focusing resulting in one or two dominant Ôpoint-likeÕ interactions and strong local deformation of the absorption surface, in contrast to the more Ôplanar-likeÕ behavior of the ideal ignition pulse. Temporal effects in moving from sub-picosecond interactions to 15-20 ps interactions, as well as from Gaussian profiles to flat-top, are beginning to be addressed through improved diagnostic measurements with ps or sub-ps resolution \cite{Ping12, Nilson12}, and through multi-ps simulation studies examining the evolution of the quantities such as the absorption efficiency, density profile, and electron distribution in time \cite{Kemp12}. Finally, one must consider whether the dominant physical processes governing the laser absorption and electron acceleration are the same in both cases. For example, effects such as ionization dynamics and the evolution of the plasma resistivity through the `warm dense' state can have a strong influence on absorption and transport in current experiments but play only a transitory role at ignition-scale. To have confidence in the extrapolation of present-day validation experiments to ignition-scale one must ensure that the dominant physical processes are similar, and if not, then to design experiments that provide better surrogacy to ignition-scale.

\section{Conclusions}
\label{sec:conclusions}

The physics of laser plasma interaction for fast ignition is an area of active research, involving the most powerful sub-nanosecond laser pulses available in experiments today,  state-of-the-art computer simulations in combination with basic kinetic theory. We have reviewed the recent literature to discuss progress in this field. We have identified the basic interaction mechanisms that are responsible for electron acceleration in regions of plasma at (1) sub-critical density, (2) near a steep density gradient, and (3) in an intermediate region. In a realistic scenario, where the laser light leads to strong modulations of the critical density surface, the distinction between these idealized cases becomes difficult. We characterize the laser-driven electron source in an idealized case of an initially flat, steep interface and discuss the influence of preformed plasma and the cone geometry, which plays an important role in terms of the cone-guided approach to fast ignition. It has been shown that the preformed plasma leads to a more energetic electron spectrum, and that the cone geometry typically leads to an enhanced preplasma scale length. 

Numerical methods for the full-scale modeling of fast ignition experiments have also seen enormous progress recently. We discuss the fundamental limitations of kinetic modeling and efforts to overcome these. In connection with these efforts we review several approaches that intend to combine a three-dimensional kinetic description of intense laser-plasma interaction on the picosecond time scale with a model of electron transport in dense matter, and describe the difficulties that each of these approaches face.

The surrogacy of current short-pulse laser experiments for a full-scale fast ignition experiment is also discussed. At full scale, fast-ignition laser pulses will need to be at least ten times more powerful than what is currently available. While present day experiments cannot be directly scaled, we discuss different ways in which they can be used to study physics issues and benchmark codes involved in designing fast ignition experiments, and the diagnostics systems that will be responsible for the tuning of future experiments.

In order to study laser plasma interaction under fast-ignition relevant conditions experimentally, our results show that laser systems should meet at least four requirements: (i) laser pulses need to be more than 1ps long, to allow for a realistic hydrodynamic plasma expansion. The early time interaction (0.5 -- 1 ps), where the plasma-vacuum interface is well defined \cite{May11}, can be significantly different from the late time interaction, where a low-density plasma shelf is generated and the laser interacts both with underdense and overdense plasma \cite{Kemp12}; (ii) laser pulses should have a spot size radius of tens of micrometers, so that the interaction in the center of the laser spot dominates over edge effects. As described above, edge effects enhance the overall electron beam divergence, compared to what we expect in a fast ignition-scale pulse \cite{Kemp12}. Moreover, intense laser pulses with narrow spot sizes ($< 10 \mu$m radius) have been shown to create a channel in the overdense plasma and lead to strong self-focusing \cite{Fiuza11}. This causes an increase of the laser intensity and therefore of the fast electron energy, which is deleterious for fast ignition, and it leads to an unstable directionality of the fast electrons due to the non-linear channel formation process \cite{Fiuza13}; (iii) laser intensities in excess of $3.5\times10^{19}$W/cm$^2$ at $1\mu$m laser wavelength, corresponding to a normalized laser amplitude $a_0=5$, in order to access the relativistic electron acceleration regime for the generation of an MeV electron population. Recent theoretical electron transport studies show that a fast electron beam energy in excess of 100 kJ near the point where the laser pulse is absorbed may be needed to achieve ignition \cite{Strozzi12, Bellei13}. This means that, for example, for a 20 ps ignition laser, 20 $\mu$m spot size, peak intensities in excess of $4 \times 10^{20}$ Wcm$^{-2}$ would be needed for full ignition scale interaction. However, our scaling arguments presented above indicate that one can use lower intensities for surrogate experiments; (iv) an energy contrast of around $10^6$ or greater is necessary to prevent significant pre-plasma formation ~\cite{MacPhee10}. Based on the considerations above, we find that current generation kilojoule-class laser facilities such as Omega-EP, delivering pulses with $t >1$ ps duration, $r > 20 \mu$m spot radius, and $I > 3.5\times10^{19}$ W/cm$^2$, can start to experimentally access the fast-ignition relevant interaction regime.

\section*{Acknowledgments}
This work was performed under the auspices of the U.S. Department of Energy by Lawrence Livermore National Laboratory under Contract DE-AC52-07NA27344. AJK and FF (Lead Coordinators of this paper) would like to thank all contributors for their efforts. AJK is supported by a DOE Early Career Award, FF is supported by the LLNL Lawrence Fellowship, WBM is supported by the DOE under Fusion Science Center through a University of Rochester subcontract No. 415025-G and under DE-FG52-09NA29552 and DE-NA0001833, and LOS is supported by the European Research Council (Accelerates ERC - 2010 - AdG Grant 267841). Computing support for this work came from the LLNL Institutional Computing Grand Challenge program and from PRACE on Juqueen based in Germany. Simulations were performed at LC-Sierra (LLNL), the Juqueen (FZ Julich, Germany), Hoffman cluster (UCLA), and IST cluster (Lisbon, Portugal).
\section*{References}

\bibliographystyle{unsrt}
\bibliography{LPI_review}

\newpage

\end{document}